\documentstyle[12pt,epsf]{article}                                 
\input{epsf}

\begin{document}

\newcommand {\bea}{\begin{eqnarray}}
\newcommand {\eea}{\end{eqnarray}}
\newcommand {\be}{\begin{equation}}
\newcommand {\ee}{\end{equation}}

\newcommand{\pref}[1]{(\ref{#1})}

\def\IR{{\hbox{{\rm I}\kern-.2em\hbox{\rm R}}}}
\def\IH{{\hbox{{\rm I}\kern-.2em\hbox{\rm H}}}}
\def\IC{{\ \hbox{{\rm I}\kern-.6em\hbox{\bf C}}}}
\def\IZ{{\hbox{{\rm Z}\kern-.4em\hbox{\rm Z}}}}

%
%
\def\journal#1&#2(#3){\unskip, \sl #1\ \bf #2 \rm(19#3) }
\def\andjournal#1&#2(#3){\sl #1~\bf #2 \rm (19#3) }
\def\nextline{\hfil\break}

\def\ie{{\it i.e.}}
\def\eg{{\it e.g.}}
\def\cf{{\it c.f.}}
\def\etal{{\it et.al.}}
\def\etc{{\it etc.}}

\def\sst{\scriptscriptstyle}
\def\tst#1{{\textstyle #1}}
\def\frac#1#2{{#1\over#2}}
\def\coeff#1#2{{\textstyle{#1\over #2}}}
\def\half{\frac12}
\def\hf{{\textstyle\half}}
\def\d{\partial}

\def\inbar{\,\vrule height1.5ex width.4pt depth0pt}
\def\IR{\relax{\rm I\kern-.18em R}}
\def\IC{\relax\hbox{$\inbar\kern-.3em{\rm C}$}}
\def\IH{\relax{\rm I\kern-.18em H}}
\def\IP{\relax{\rm I\kern-.18em P}}
\def\Z{{\bf Z}}
\def\One{{1\hskip -3pt {\rm l}}}
\def\nth{$n^{\rm th}$}
%
%
\def\np#1#2#3{Nucl. Phys. {\bf B#1} (#2) #3}
\def\npb#1#2#3{Nucl. Phys. {\bf B#1} (#2) #3}
\def\pl#1#2#3{Phys. Lett. {\bf #1B} (#2) #3}
\def\plb#1#2#3{Phys. Lett. {\bf #1B} (#2) #3}
\def\prl#1#2#3{Phys. Rev. Lett. {\bf #1} (#2) #3}
\def\physrev#1#2#3{Phys. Rev. {\bf D#1} (#2) #3}
\def\prd#1#2#3{Phys. Rev. {\bf D#1} (#2) #3}
\def\annphys#1#2#3{Ann. Phys. {\bf #1} (#2) #3}
\def\prep#1#2#3{Phys. Rep. {\bf #1} (#2) #3}
\def\rmp#1#2#3{Rev. Mod. Phys. {\bf #1} (#2) #3}
\def\cmp#1#2#3{Comm. Math. Phys. {\bf #1} (#2) #3}
\def\cqg#1#2#3{Class. Quant. Grav. {\bf #1} (#2) #3}
\def\mpl#1#2#3{Mod. Phys. Lett. {\bf #1} (#2) #3}
\def\ijmp#1#2#3{Int. J. Mod. Phys. {\bf A#1} (#2) #3}
\def\jmp#1#2#3{J. Math. Phys. {\bf #1} (#2) #3}
\catcode`\@=11
\def\slash#1{\mathord{\mathpalette\c@ncel{#1}}}
\overfullrule=0pt
\def\AA{{\cal A}}
\def\BB{{\cal B}}
\def\CC{{\cal C}}
\def\DD{{\cal D}}
\def\EE{{\cal E}}
\def\FF{{\cal F}}
\def\GG{{\cal G}}
\def\HH{{\cal H}}
\def\II{{\cal I}}
\def\JJ{{\cal J}}
\def\KK{{\cal K}}
\def\LL{{\cal L}}
\def\MM{{\cal M}}
\def\NN{{\cal N}}
\def\OO{{\cal O}}
\def\PP{{\cal P}}
\def\QQ{{\cal Q}}
\def\RR{{\cal R}}
\def\SS{{\cal S}}
\def\TT{{\cal T}}
\def\UU{{\cal U}}
\def\VV{{\cal V}}
\def\WW{{\cal W}}
\def\XX{{\cal X}}
\def\YY{{\cal Y}}
\def\ZZ{{\cal Z}}
\def\lam{\lambda}
\def\eps{\epsilon}
\def\vareps{\varepsilon}
\def\underrel#1\over#2{\mathrel{\mathop{\kern\z@#1}\limits_{#2}}}
\def\lapprox{{\underrel{\scriptstyle<}\over\sim}}
\def\lessapprox{{\buildrel{<}\over{\scriptstyle\sim}}}
\catcode`\@=12

\def\sdtimes{\mathbin{\hbox{\hskip2pt\vrule height 5.5pt depth -.3pt width
.4pt \hskip-2pt$\times$}}}
\def\bra#1{\left\langle #1\right|}
\def\ket#1{\left| #1\right\rangle}
\def\vev#1{\left\langle #1 \right\rangle}
\def\det{{\rm det}}
\def\tr{{\rm tr}}
\def\mod{{\rm mod}}
\def\sinh{{\rm sinh}}
\def\cosh{{\rm cosh}}
\def\sgn{{\rm sgn}}
\def\det{{\rm det}}
\def\exp{{\rm exp}}
\def\sh{{\rm sh}}
\def\ch{{\rm ch}}

\def\lpl{\ell_{{\rm pl}}}
\def\lstr{l_s}
\def\str{{\sst\rm str}}
\def\bps{{\rm BPS}}
\def\ads{{\rm AdS}}
\def\inst{{\rm inst}}
\def\sltwoz{SL(2,\Z)}
\def\rhot{{\tilde\rho}}
\def\taut{{\tilde\tau}}
\def\Ptil{{\tilde P}}
\def\Ttil{{\widetilde \TT}}
\def\Jtil{{\widetilde \JJ}}
\def\Jtilbar{{\overline{\widetilde \JJ}}}
\def\extrat{{\tilde T}}
\def\ellt{{\tilde\ell}}
\def\gs{g_{s}}
\def\gsix{g_6}
\def\nufour{\nu_4}
\def\adot{\dot a}
\def\Ytil{{X}}
\def\lamt{{\psi}}
\def\kth{$k^{\rm th}$}
\def\symn{{\sl Sym}^N}
\def\ghat{{\hat G}}
\def\hq{{\HH_{\vec q}}}
\def\eee{\varepsilon}


\begin{titlepage}
\rightline{EFI-99-18}

\rightline{hep-th/9905064}

\vskip 3cm
\centerline{\Large{\bf U(1) Charges and Moduli in the D1-D5 System}}

\vskip 2cm
\centerline{
Finn Larsen\footnote{\texttt{flarsen@theory.uchicago.edu}}~~ {\it and} ~~
Emil Martinec\footnote{\texttt{ejm@theory.uchicago.edu}}}
\vskip 12pt
\centerline{\sl Enrico Fermi Inst. and Dept. of Physics}
\centerline{\sl University of Chicago}
\centerline{\sl 5640 S. Ellis Ave., Chicago, IL 60637, USA}

\vskip 2cm
\begin{abstract}
The decoupling limit of the D1-D5 system compactified on
$T^4\times S^1$ has a rich spectrum of U(1) charged
excitations.  Even though these states are not BPS in the limit,
BPS considerations determine the mass and the semiclassical
entropy for a given charge vector. The dependence of the mass 
formula on the compactification moduli
situates the symmetric orbifold ${\sl Sym}^N(T^4)\times\tilde T^4$
conformal field theory in the moduli space.  
A detailed analysis of the global identifications
of the moduli space yields a picture of multiple weak-coupling limits
-- one for each factorization of $N$ into D1 and D5 charges
$d_1$ and $d_5=N/d_1$ -- joined through regions of strong coupling
in the CFT moduli space.
\end{abstract}

\end{titlepage}

\newpage
\setcounter{page}{1}


\section{Introduction and Summary\label{intro}}

The D1-D5 system is a touchstone of recent progress in 
string theory.  It underlies the first reliable counting
of black hole microstates~\cite{Strominger:1996sh}; 
as well as precise calculations of their cross-sections 
for absorption and emission of low-energy quanta, in the context of 
a unitary quantum-mechanical theory 
(for review see {\it e.g.}~\cite{Maldacena:1997tm,Peet:1997es}).
It is a prime example of
the duality between gravitational and non-gravitational
dynamics~\cite{Banks:1997vh,Maldacena:1997re,gkp,Witten:1998qj}
in certain scaling limits.

When the D1-D5 system is compactified on $X^4=T^4$ or K3
in the directions on the D5-brane that are
transverse to the D1-brane, and on a large circle in the common direction,
the low-energy dynamics in the appropriate
scaling limit appears to be generically described by a sigma
model on the moduli space of instantons $\MM_\inst$ in the D5-brane
gauge theory.%
\footnote{The ADHM equations defining this moduli space
can be suitably generalized to include fivebrane number equal to one;
the resulting variety is called the Hilbert scheme 
(\cf~\cite{Dijkgraaf:1998gf}).}
It has been proposed that this sigma model might be effectively described 
as a blowup of the symmetric product orbifold $\symn(X^4)$.

When $X^4=T^4$, there is a large class of 
of U(1) charged excitations.  Although these
excitations are not BPS after taking the scaling limit,
they {\it are} BPS states beforehand; moreover, their 
energies remain finite in the limiting theory.
BPS considerations thus determine a lower bound on the masses
of charged states, as well as the semiclassical entropy
of states satisfying the bound; the evaluation of these
quantities constitutes some of our main results. 
Their determination is not solely attributable 
to the fact that these charges couple to a U(1) current algebra in 
the limiting theory.  

We begin in section \ref{scalimit} 
with a specification of the scaling limit
itself, as a limit of the BPS mass formula where certain
charges are taken to describe a ``heavy'' brane background,
while others remain ``light''.
According to the Maldacena conjecture~\cite{Maldacena:1997re},
the dynamics of the heavy brane background is dual to
string theory on $AdS_3\times S^3\times X^4$,
where the anti-de Sitter radius in units of the fundamental
string scale is $R_{\ads}=\lstr(\gsix^2q_1q_5)^{\frac14}$ ($\gsix$
is the six-dimensional string coupling), and the characteristic
proper size of $X^4$ is $\lstr(q_1/q_5)^{\frac14}$.
We determine how the energy of the light U(1)-charged excitations 
on $X^4=T^4$ depend on the radii of a rectangular four-torus,
with all other moduli set to zero.
The resulting mass formula depends separately on the background 
one-brane and five-brane charges $q_1$ and $q_5$ 
(and not just the product $q_1q_5$), 
due to the different ways the proper size of the
four-torus affects the various charged objects~\cite{Kutasov:1999xu}.
This indicates that the theory in some way distinguishes
backgrounds with different one-brane and five-brane charges.

This leads us in section \ref{massformula}
to a more detailed investigation of the 
dependence of the BPS mass formula on all the 
$SO(5,5)/SO(5)\times SO(5)$
moduli of compactification on $T^4$.
We begin with an analysis of the heavy brane background,
determining the tension of objects wrapping the large circle.
The minimization of this tension fixes five of the moduli
in terms of the rest~\cite{Ferrara:1995ih,Ferrara:1996dd}, reducing the 
local geometry of the moduli space
to $\KK=SO(5,4)/SO(5)\times SO(4)$.  We give a set of explicit 
and general formulae for the fixed scalars.  

For generic moduli, the D1- and D5-branes are bound together
by an amount determined from the tension formula.
The bound state dynamics is the Higgs branch of
the low-energy D1-D5 gauge theory, reducing in the infrared
to the aforementioned sigma model on instanton moduli space $\MM_\inst$;
the singularities of the instanton moduli space 
are regularized at generic
points in the Teichmuller space $\KK$ by 
nonzero antisymmetric tensor backgrounds 
(including the RR scalar) 
in the ambient spacetime~\cite{Seiberg:1999xz,Dijkgraaf:1998gf}.  
The binding energy of the brane background
vanishes along certain codimension four
subspaces of $\KK$.%
\footnote{With periodic (Ramond) boundary conditions on fermions.
As discussed in~\cite{Seiberg:1999xz}, there is a finite gap
between the two branches in the presence of
antiperiodic (NS) boundary conditions on the fermions.} 
These are the domains where the D1-D5 bound states 
can separate into subsystems.
Coulomb branches (more precisely mixed Coulomb-Higgs branches) 
of the D1-D5 gauge dynamics describe configurations where these
subsystems are separated in directions transverse to the 
fivebranes.  It is the appearance of these new branches that
causes the spacetime CFT to become singular.%
\footnote{It is often said that the Coulomb and Higgs branches
of the field space decouple in the low-energy limit of the
brane dynamics; however, since the effective field theory 
that describes the singularity of the Higgs 
branch~\cite{Maldacena:1999uz,Seiberg:1999xz} is written
in terms of a vector multiplet describing the transverse separation of
branes, one might say that the Higgs branch is joined smoothly
onto the near-horizon region of the Coulomb branch.
Similar phenomena occur in the parametrization of the large-N
solution of the ADHM equations~\cite{Dorey:1999pd}, where the scale
size of instantons inside a stack of D3-branes 
is isomorphic to the transverse
separation of D-instantons from the D3-branes, provided that 
separation is scaled to remain finite in the Maldacena limit.
We thank A. Strominger for discussions of this issue.} 
We apply the tension formula to determine the singular locus
in the moduli space $\KK$ for given background brane charges
$d_1$ and $d_5$ (including the case where either of the charges
is equal to one).

Having dealt with the heavy background for generic moduli,
we proceed to an analysis of the light U(1)-charged excitations
and their dependence on the moduli.  This exercise provides
a great deal of robust, non-topological data that can be
compared to particular conformal field theories such as the
symmetric orbifold.  Later, in section \ref{symprod}, we use this data
to locate the subspace of the moduli space described
by the symmetric orbifold.

In maximal supergravity on $T^5$, 1/8-BPS states have
a finite entropy.  Usually, the charges participating 
in this entropy are taken to be the background D1- and D5-brane
charges themselves, together with momentum along their
common direction.  However, the finite entropy will persist
even when the responsible charges are ``misaligned''
with the large circle in the scaling procedure of section \ref{scalimit} 
-- for example, a string could carry both winding and momentum
on the small $T^4$.
This finite entropy of states carrying combinations
of U(1) charges is calculated in section \ref{entropy}.

An important part of the structure of moduli spaces of 
toroidal compactifications is their group of global identifications.
These are discrete transformations on the charge lattice
and moduli that leave the spectrum invariant.
With the full moduli dependence of the charged spectrum in hand,
we are poised for an analysis of these identifications in section 
\ref{globalident}.
In the presence of the ``heavy'' brane background, the full
$SO(5,5;\Z)$ duality group of string theory compactified on $T^5$
is reduced to the subgroup $\HH_{\vec q}$ that preserves 
the charge vector of the background.
Transformations not in this subgroup do not preserve the fixed
scalar conditions or the spectrum of U(1) charged excitations.
Nevertheless, the theory is {\it covariant} under such transformations;
one can transform any given background charges $(q_1,q_5)$
into $(N=q_1q_5,1)$, as long as one transforms the moduli
(including the fixed scalars) at the same time.
Therefore, the moduli space for charges $(N,1)$ will be 
continuously connected to regions having a natural interpretation
in terms of charges $(q_1,q_5)$.  These are simply
two different interpretations of the {\it same} 
spacetime conformal field theory.
We call the charge vector $(N,1)$ the {\it canonical background}.

It is easy to find transformations from a given charge
vector to the canonical charge vector inside an $SO(2,2;\Z)$
subgroup of the duality group; a convenient choice is
the $SO(2,2;\Z)$ subgroup acting on the moduli that are scalar
or pseudoscalar on the $T^4$.  
The map from the background charges $(q_1,q_5)$
to the canonical charges transforms the singular
locus in moduli space to a well-defined region in the fundamental
domain of the moduli in the canonical background.
The interesting part of the residual duality group $\HH_{\vec q}$
turns out to be a certain `diagonal' $\Gamma_0(N)$
inside this $SO(2,2;\Z)$.%
\footnote{$\Gamma_0(N)$ is the subgroup of $\sltwoz$ defined by
matrices $({a~b\atop c~d})$, $ad-bc=1$, $c\equiv 0$~mod~$N$. }
We show how the important structure of the fundamental domain 
of the moduli space of the spacetime CFT
inside the Teichmuller space $\KK=SO(5,4)/SO(5)\times SO(4)$
can be projected onto the fundamental domain of $\Gamma_0(N)$
acting on $\tau=\chi+i/\gs$ (here $\chi$ is the RR scalar, $\gs$
the string coupling of type IIB string theory).
The rational cusps of the fundamental
domain have the interpretation as coming from D1-D5 systems
having different partitions of $N$
into one-brane and fivebrane numbers $(d_1,d_5=N/d_1)$.
The singular locus in the CFT moduli space, where the system
is unstable to decay by emitting D1- and D5-branes,
is located as a set of circular arcs inside this fundamental domain
passing between conjugate cusps.  A satisfying picture of
the CFT moduli space emerges, whereby different weakly-coupled
regions of the CFT, describing the spacetime physics of different
numbers of D1- and D5-branes, are connected through regions
of strong coupling.

One of the goals of our investigation was to pin down the
relation between the symmetric orbifold
$\symn(X^4)$ and the D1-D5 spacetime conformal field theory.
In section \ref{symprod}, we match the structure of a somewhat
modified symmetric orbifold ${\sl Sym}^N(T^4)\sdtimes\extrat^4$
to that of the spacetime CFT. The extra $\extrat^4$ is needed to 
represent the full spectrum of sixteen $U(1)$ charges. 
The U(1) currents of this orbifold naturally 
split into two sets: Eight (four left-moving and four right-moving)
of ``level'' $N$ from the symmetric product,%
\footnote{By the level of a U(1) current algebra, we mean the
coefficient of the double pole in the current-current two-point function,
when the currents are canonically normalized 
such that the momentum and winding charges 
are $p_i/r_i$ and $w^ir_i$ as in equation \pref{eq:er0} below.}
and eight more of level one from the extra $\extrat^4$.
This suggests that the orbifold naturally describes a region
in the cusp of the fundamental domain corresponding
to the canonical charges $(N,1)$; we indeed find that
all known data are consistent with this proposal.
This data includes the precise spectrum
of BPS states (including the upper cutoff on R-charges),
which matches expectations from 
duality~\cite{Vafa:1996zh,Maldacena:1998bw,Martinec:1999sa};
the qualitative growth of the full (non-BPS) spectrum~\cite{Banks:1998dd};
the structure of the moduli taking one away from the
orbifold locus; and the U(1) mass formula.

During the course of our investigations, we learned of related 
work~\cite{Maldacena:1999bp} on the entropy of U(1) charged
excitations; and on the global structure of the moduli 
space~\cite{Seiberg:1999xz}.


\section{The Setting\label{scalimit}}

Consider type IIB string theory toroidally compactified to five 
dimensions. This theory contains 27 $U(1)$ charges: 5 KK-momenta, 
5 fundamental strings,  1 fully wrapped NS5-brane, ${5\choose 1} =5$ 
D1-branes (D-strings), ${5\choose 3}=10$ D3-branes, and ${5\choose 5}=1$ 
D5 brane. The $U(1)$ charges transform in the fundamental $\bf 27$ of the 
$E_{6(6)}$ duality group. The toroidal compactification is characterized 
by $42$ moduli that parametrize the coset space $E_{6(6)}/USp(8)$. In 
this section we focus on the moduli that are given by the radii of a 
rectangular torus, as well as the type IIB coupling constant.

\paragraph{The Scaling Limit:} 
The microscopic description of black 
holes is based on a field theory that 
does not contain gravity. This theory can be isolated from the full string 
theory by taking a suitable limit. The limit leaves one of the radii, 
say $R_5$, much larger than the other four radii 
$R_i$, $i=1,2,3,4$.  More precisely, we take
$l_s\to 0$ with $R_i \propto l_s $ ($i=1,2,3,4$) and $R_5$ fixed,
while keeping the integer quantized 
values of the $U(1)$ charges fixed. 
The scale of energy in the system is set by $R_5^{-1}$.
The effect of this limit on the
various $U(1)$ charges can be judged by considering the mass of 
singly charged constituent branes:\footnote{The precise definition
of the string scale is $l_s\equiv \sqrt{\alpha^\prime}$.}
\bea
 M_{F1} &=& R_i/l^2_s\nonumber\\
 M_{KK} &=& 1/R_i \nonumber\\
 M_{NS5} &=& R_1 R_2 R_3 R_4 R_5/l^6_s g^2_s\nonumber\\
 M_{D1} &=& R_i/l^2_s g_s\label{masses}\\
 M_{D3} &=& R_i R_j R_k/l^4_s g_s\nonumber\\
 M_{D5} &=& R_1 R_2 R_3 R_4 R_5/l^6_s g_s\ .\nonumber
\eea
The 10 branes that wrap $R_5$ have masses that scale as $l^{-2}_s$. 
These are the NS5/D5, the F1/D1 along $R_5$, and the ${4\choose 2}=6$ 
D3 branes with one dimension along $R_5$. The branes that do not
wrap $R_5$ and the KK-momenta within the small $T^4$ have masses that 
scale as $l^{-1}_s$; there are 16 such charges. Finally the KK momentum 
along $R_5$ gives a mass scaling as $\lstr^0$.

In the decoupling limit four dimensions are taken small, so it is 
natural to interpret results in terms of the remaining six dimensions
even though one of these is actually compact (with radius $R_5$). 
The $10,16,1$ charges with masses that scale as $l^{-2}_s$, $l^{-1}_s$
and $l^{0}_s$ are tensors, vectors, and scalars from this six dimensional
point of view. They transform in the vector ({\bf 10}), 
spinor ({\bf 16}) and scalar ({\bf 1}) of the $SO(5,5)$ duality group 
of the six-dimensional theory. 
The most massive excitations are those that correspond to charges of 
the 10 tensor fields. The strategy is to consider a background created
by these ``heavy'' charges and then consider the remaining ``light''
charges as excitations in the resulting theory.

\paragraph{The Tensor Background:} 
Accordingly, first consider the tensor charges. 
Regular black holes correspond to configurations 
where these preserve 1/4 of the supersymmetry. 
Simple choices of excited charges are the D1 and the D5, or the F1 and 
the NS5 (these and other choices are equivalent under the
action of the $SO(5,5;\Z)$ U-duality subgroup that preserves
the distinction between the small and large circles).
We consider the F1/NS5 system without 
loss of generality.  The mass of this brane background is
\be
  M =\frac{R_5}{\lstr^2}\Bigl( q_1+\frac{v_4}{\gs^2}q_5\Bigr)\ ,
\label{eq:f1f5mass}
\ee
where $v_4=R_1R_2R_3R_4/\lstr^4$.  Due to the attractor mechanism
(see below), in the near-horizon geometry the six-dimensional
string coupling $\gsix^2=\gs^2/v_4$
is drawn to $\gsix^2=q_5/q_1$, the value that minimizes the mass
of the brane background in 6d Planck units.

\paragraph{The Vector Excitations:} 
Next consider the charged 
excitations about this background. It is 
instructive to begin the discussion with a simple example. Consider 
the F1/NS5 system with momentum along the F1 on a rectangular
$T^5$ with vanishing parity-odd moduli. In general the F1 is not aligned with
any of the coordinate axes, but the mass of the configuration is 
nevertheless the sum of constituent masses
\be
 l_s M = {v_4 r_5\over g^2_s} n_{5} +   
\sqrt{\sum_{a=1}^5 (w_a^{F1} r^a+ {p^a\over r^a})^2 }\ ,
\ee
where $r_a=R_a/l_s$ for $a=1,\ldots,5$ (so $v_4\equiv r_1 r_2 r_3 r_4$). 
In the scaling limit the formula becomes
\be
l_s M = {v_4 r_5\over g^2_s} n_{5} +  f_{1} r^5 + 
{p^5\over r^5} + {1\over 2q_{F1} r^5}
\sum_{i=1}^4 \left(w_i^{F1} r^i+ {p^i\over r^i}\right)^2\ ,
\ee
where terms that vanish in the limit were omitted and 
$f_{1}\equiv w^{F1}_5$. The first two terms scale as $l^{-2}_s$, 
as expected for tensor field backgrounds. More importantly, 
{\it there are no terms that scale as} $l^{-1}_s$; the energy attributable 
to the vector excitations is of order $l_s^0$. In other words, the 
vector charges contribute much less to the energy in the environment created 
by the tensor fields than they would in isolation. 

After the remaining vector charges are taken into account, the energy of 
the vector excitations becomes
\be
  R_5 E = 
	p_5+{1\over 2f_1}\sum_{i=1}^4\Bigl({p_i\over r_i}+w^i_{F1}r_i\Bigr)^2
	+{1\over 2n_5 }\sum_{i=1}^4\Bigl({w^{D3}_i\sqrt{v_4}\over r_i}
	+w^i_{D1}\frac{r_i}{\sqrt v_4}\Bigr)^2\ ,
\label{eq:er0}
\ee
where $w^{D3}_{i}\equiv {1\over 6}\epsilon_{ijkl}w_{D3}^{jkl}$ are wrapping 
numbers of D3 branes that wrap the four-torus. This formula
is derived in the Appendix. It can be motivated by noting that
the two last terms in (\ref{eq:er0}) are related by the symmetries
of the theory; indeed $ST_{1234}S$ performs the interchanges 
$f_1\leftrightarrow n_5$, $p_i\leftrightarrow w_{D1}^i$, and 
$w_{F1}^i\leftrightarrow w_{D3\,i}$.

The formula \pref{eq:er0} has the same spectrum as
a ``one-brane'' current algebra at level $f_1$ on a torus of radii $r_i$,
together with a ``five-brane'' current algebra at level $n_5$
on a torus of radii $r_i/\sqrt{v_4}$.  This leads to a 
puzzle~\cite{Kutasov:1998zh}: the proposed dual symmetric orbifold CFT 
has a diagonal $U(1)^4$ current algebra in the untwisted
sector of level $N=f_1 n_5$,
and so would appear at best to describe $f_1=N$, $n_5=1$.
We will see in section \ref{globalident}
that this apparent difficulty is an artifact
of the particular subspace of moduli taken into account
in \pref{eq:er0}.  At generic points in moduli space, the
mass formula does not neatly separate into a ``one-brane term''
and a ``five-brane term''. 


BPS arguments guarantee that the masses given in equation \pref{eq:er0} 
are exact, that is, there are no perturbative or non-perturbative 
corrections. Nevertheless, {\it the excitations are not BPS in the 
decoupled theory}. It is the complete set of supersymmetries in the 
original string theory that ensures the exactness and some of these 
are realized nonlinearly in the decoupled theory. One of the 
motivations for considering the $U(1)$ charges is that these are 
well-behaved non-BPS excitations.

In this sense, our situation is directly parallel to
that of matrix theory, where parts of the eleven-dimensional
supersymmetry algebra become nonlinearly realized in the
limiting process that defines the construction, and certain BPS
charges (in that case, the transverse fivebrane charge)
decouple from the supersymmetry algebra~\cite{Banks:1997nn}.  
This property is a generic feature of the decoupling limit.

The mass formula \pref{eq:er0} 
gives the energy of the lowest state with the specified charges.
More generally this formula can be interpreted as a lower bound on the
energy in the superselection sector with the specified charges. In this
way the considerations of this section are relevant at any finite
energy in the decoupled theory. 

The discussion in this section assumed toroidal compactification.
It is straightforward to replace the small four-torus with a $K3$ and 
scale the volume of the $K3$ in the same way.
Then the ``heavy'' charges are carried by $26$ tensor fields 
in the 6d theory that transform in the fundamental representation
of the duality group $SO(5,21)$.  
However, there are no 6d vectors and so there
is less to say about the structure.%
\footnote{Another puzzle, raised in~\cite{Martinec:1999sa},
concerns whether a 1+1d CFT can describe the limits
in the K3 moduli space where a vanishing cycle on the K3 appears.
At these points, a tensionless string appears in the spectrum 
of IIB string theory; where is this string in the spectrum
of the CFT?  First of all, tensor charges (for either
$T^4$ or K3) are ``invisible'' in the sense that they are part of
the specification of the CFT background rather than an
excitation of the CFT itself.  Therefore, the appearance of a light
string in the spectrum (which need not be one of those
in the brane background) means that the target space of the CFT
is becoming strongly curved and/or of small volume
in units of that string's tension, and therefore at least
part of the CFT is becoming strongly coupled.  The decoupling
limit assumed that all the background strings had finite tension
in the limit, and this assumption is breaking down.
Using duality transformations of the type explored below in
section \ref{globalident}, 
one can relate these limits to decompactification
limits in some other duality frame.}

\section{Masses and Moduli}
\label{massformula}

The energy carried by $U(1)$ charged excitations depends sensitively 
on the moduli of the background. The purpose of this section is to 
present a mass formula that expresses the dependence on general 
$SO(5,4)/SO(5)\times SO(4)$ moduli. 

\paragraph{The Tensor Fields:}
Consider first the heavy background, {\it i.e.} the 6d tensor fields,
and further assume that the parity-odd moduli vanish. In this
restricted case the square of the mass is the obvious norm of the
$SO(5,5)$ charge vector. We write the result as
\bea
{{\tilde G}_6 M^2_5\over R_5^2} &=& {g^2_s\over v_4} 
(f_{1} + {v_4\over g^2_s}n_{5})^2 
+ {1\over v_4}(d_{1} + v_4 d_{5})^2 +  \nonumber \\
&+& {1\over 2v_4} D^{ij5}G_{ik} G_{jl}D^{kl5} +
{1\over 4}\epsilon_{ijkl} D^{ij5}D^{kl5}\ ,
\label{eq:prelm}
\eea
where $D^{ij5}$ denote the numbers of D3-branes with one dimension
wrapping $R^5$, and 
$G_{ij}$ is the string metric ({\it e.g.}
$G_{11}=r^2_1$ on a square torus). 
The normalization factor
\be
{\tilde G}_6 = {l_s^2 g^2_s\over v_4} = {8\over (2\pi)^2}G_6~,
\ee
is convenient because it is invariant under $SO(5,5)$ duality
transformations. 
The normalization of the various 
terms in (\ref{eq:prelm}) can be verified by writing the right hand 
side as the sum of five perfect squares; the first is the square of 
(\ref{eq:f1f5mass}), and the others follow similarly by comparison with 
(\ref{masses}).

The next step is to include the parity-odd moduli in the background. 
It turns out that the general effect of these fields can be summarized 
through the substitution rules
\bea
f_{1} \to  {\tilde f}_{1}&=&
f_{1}  + \chi {\tilde d}_{1} 
- (A_4 - {1\over 8}\epsilon^{ijkl}B_{ij}C_{kl})d_{5} -
{1\over 8}\epsilon^{ijkl}C_{ij}C_{kl}n_{5}  
-{1\over 2}C_{ij} D^{5ij} \nonumber \\
d_{1} \to {\tilde d}_{1} &=& d_{1} 
+ (A_4+{1\over 8}\epsilon^{ijkl}B_{ij}C_{kl}) n_{5} 
- {1\over 8}\epsilon^{ijkl}B_{ij}B_{kl}d_{5}+{1\over 2}B_{ij} D^{5ij}
\label{eq:bshift}\\
D^{ij5} \to  {\tilde D}^{ij5}&=&
D^{ij5} - {1\over 2}\epsilon^{ijkl}B_{kl}d_{5} + 
{1\over 2}\epsilon^{ijkl}C_{kl}n_{5}  \nonumber \\
n_{5}\to  \tilde{n}_{5}&=&n_{5}\nonumber \\
d_{5}\to  \tilde{d}_{5}&=&d_{5} -\chi n_5 \ .\nonumber 
\eea
The physical interpretation of these shifts is that the parity-odd moduli 
induce charges in addition to the those that are present in the background. 
This effect is well-known from the anomaly-induced charges on D-brane 
world-volumes~\cite{Douglas:1995bn} 
and, simpler yet, from the shift in the momentum of the perturbative 
string in the presence of a NS B-field. The full substitution rules 
(\ref{eq:bshift}) can be derived using the methods 
reviewed in~\cite{Obers:1998rn}.

\paragraph{The Fixed Scalars:}
The classical solutions that correspond to a given charge assignment
have the property that, in the near horizon region, some of the background
moduli are attracted to values which are independent 
of the moduli in the asymptotically flat region, and which 
are determined exclusively by the charge 
vector~\cite{Ferrara:1995ih,Ferrara:1996dd}. From the perspective of 
the theory without gravity these scalars are {\it not} moduli because 
they cannot be varied; they are known as fixed scalars. Their values can 
be determined in general by minimizing the mass formula (\ref{eq:prelm}) 
with the substitutions (\ref{eq:bshift}).

It is instructive to carry out the details in the D1/D5 case.
Then the mass formula is
\bea
  {{\tilde G}_6 M^2_5\over R^2_5}  &=& {1\over v_4}
	\left[d_{1} + (v_4-{1\over 8}\epsilon^{ijkl}B_{ij} B_{kl}) d_{5} 
	\right]^2 \nonumber \\
& & \qquad 
	+ {g_s^2\over v_4} 
\left[ \chi (d_{1}- {1\over 8}\epsilon^{ijkl}B_{ij} B_{kl})  
-  (A_4 - {1\over 8} \epsilon^{ijkl}B_{ij} C_{kl} ) d_{5} 
\right]^2 \nonumber \\
& &\qquad +  
	\left[v_4~{1\over 8} B_{ij} G^{ik}G^{jl}B_{kl}
	+{1\over 4}\epsilon^{ijkl}B_{ij}B_{kl}\right] d^2_{5}\ .
\label{eq:dtension}
\eea
The fixed scalars are determined by minimizing over moduli space. 
The result is the self-duality conditions
\bea
v_4 B_{ij} G^{ik}G^{jl} &=&  {1\over 2}B_{ij}\epsilon^{ijkl} 
\label{eq:dfixed1}\\
 v_4 \chi &= & A_4 - {1\over 8}\epsilon^{ijkl}B_{ij} C_{kl}~,
\label{eq:dfixed2}
\eea
and the fixed volume condition
\be
v_4 + {1\over 8}\epsilon^{ijkl} B_{ij} B_{kl}  
=  {d_{1}\over d_{5}}\ .
\label{eq:dfixed3}
\ee
The mass formula at the fixed scalar point is
\be
{{\tilde G}_6 M^2_5\over R_5^2} = 4 d_{1} d_{5}\ .
\label{eq:dfixm}
\ee
The free moduli that remain in the limiting theory can be chosen 
as the $16$ fields $G_{ij}+C_{ij}$ as well as the RR-scalar $\chi$ and 
the self-dual NS tensor fields, denoted $B^+$. 
The tensor mass at the 
fixed point is independent of all these 20 moduli, as it should be.

It is also interesting to consider the special case of the F1/NS5 background.
Then the mass formula is
\bea
  {{\tilde G}_6 M^2_5\over R^2_5} &=& {g^2_s\over v_4}
	\left[f_{1} - {1\over 8}\epsilon^{ijkl} C_{ij} C_{kl} n_{5}
	+ {1\over 8}\epsilon^{ijkl} B_{ij} C_{kl} \chi n_{5}
	+(\chi A_4 + {v_4\over g^2_s})n_{5} \right]^2 \nonumber \\
& &\qquad + {1\over v_4}
	\left[{1\over 8}\epsilon^{ijkl} B_{ij} C_{kl} + A_4 
    -\chi v_4 \right]^2 n_{5}^2 \nonumber \\
& &\qquad + \left[v_4~{1\over 8} C_{ij} G^{ik}G^{jl}C_{kl}
	+{1\over 4}\epsilon^{ijkl} C_{ij}C_{kl}\right]n^2_{5}\ .
\label{eq:ftension}
\eea 
The fixed scalar conditions become
\bea
v_4 C_{ij} G^{ik}G^{jl} &=& {1\over 2}C_{ij}\epsilon^{ijkl}
\label{eq:ffixed1} \\
v_4 \chi &= & - A_4 - {1\over 8}\epsilon^{ijkl}B_{ij} C_{kl}
\label{eq:ffixed2} \\
{f_{1}\over n_{5}} &=&
v_4 (\chi^2 + {1\over g^2_s}) + 
{1\over 8}\epsilon^{ijkl} C_{ij} C_{kl}  \ .
\label{eq:ffixed3}
\eea
As a check on our computations we have verified that the fixed
scalar equations (\ref{eq:dfixed1}-\ref{eq:dfixed3}) transform
into (\ref{eq:ffixed1}-\ref{eq:ffixed3}) under duality.
The mass formula at the fixed scalar point is similarly
the dual of (\ref{eq:dfixm})
\be
{\tilde {G}_6 M^2_5\over R_5^2} = 4 f_{1}n_{5}\ ,
\label{eq:ffixm}
\ee
as expected. In the F1/NS5 system
the free moduli can be chosen as the $16$ fields $G_{ij}+B_{ij}$ as well as 
the RR-scalar $\chi$ and the self-dual RR tensor-fields, denoted $C^+$.

\paragraph{The Singularities in Moduli Space:}
The spacetime conformal field theory is singular in a 
subspace of moduli space~\cite{Witten:1997yu}. The physical origin
of the singularity was explained 
in~\cite{Maldacena:1999uz,Seiberg:1999xz}: The 
D1/D5 system can become unstable to fragmentation into smaller 
constituents.  Decay is only possible for special values of moduli, 
which constitute the singular locus of moduli space.

The singular locus of moduli space is determined by the tensor mass formula.
For example, consider the decay of the D1/D5 brane system 
with charges $(d_1,d_5)$ into two fragments with charges 
$(d_1-d_1^\prime,d_5)$ and $(d_1^\prime,0)$, respectively 
(with $0<d_1^\prime<d_1$). The
mass of each constitutent is given by (\ref{eq:dtension}),
with the background moduli constrained by the fixed scalar conditions 
of the initial state (\ref{eq:dfixed1}-\ref{eq:dfixed3}). 
We further assume for simplicity 
that the $B^+$ vanishes and that the RR-scalar $\chi$ is small. 
Then the change in mass from initial to final state is given by
\be
{\delta M_5\over M_5} = - {\chi^2 g^2_s\over 2}
~{d_1^\prime\over 2d_1-d_1^\prime} + {\cal O}(\chi^4)\ .
\ee
Thus the decay is kinematically forbidden, when $\chi\neq 0$. 
On the other hand, when $\chi=0$ there is no barrier to decay, 
so the singular locus of moduli space includes 
the codimension four subspace given by $\chi=B^+=0$. 
The converse result is also true, up to duality:
Fragmentation is possible precisely for the moduli $\chi=B^+=0$, or 
others that are related to them by 
duality\footnote{The global subtleties related to duality are discussed 
in sec. 5.}. It is straightforward to show that other potential
decays are also forbidden unless $\chi=B^+=0$.
Note that since these results are underpinned by BPS arguments,
they are valid for all values of the charges and in particular when 
$d_5=1$. This is a necessary consistency condition for the proposal that 
the symmetric product describes the case $d_5=1$. 

In~\cite{Maldacena:1999uz,Seiberg:1999xz} a concrete picture of the 
singularity in the spacetime CFT was achieved in terms of an effective 
sigma model of long strings. This description breaks down completely for 
$d_5=1$; and for $d_5=2$ the model cannot describe the perturbation 
which lifts the degeneracy of the Coulomb and Higgs branches (it is not 
in the chiral ring of this effective theory for $d_5=2$). It would be 
interesting to find a useful description of the singularity of the 
spacetime CFT in these two special cases.

\paragraph{The Vector Fields:}
The next step is to add vector charges to the background of tensor fields, 
and then expand as specified by the scaling limit discussed in 
section \ref{scalimit}. The 
energy associated with the vectors is the total energy of the system less 
the mass of the background. 

We first consider a square torus and assume that no non-trivial moduli
are turned on. The desired mass formula follows from BPS algebra, as detailed
in the Appendix. In the case where the background is the F1/NS5 system 
the result is
\be
E R_5 = {1\over 2f_{1}}~(\vec{p} + \vec{w}_{F1})^2
 + {1\over 2n_{5}}~{1\over v_4}~(\vec{w}_{D1} + v_4\vec{w}_{D3})^2\ .
\label{eq:vecef}
\ee
The notation here and in the following is that the metric contracting 
$\vec{p}$ or $\vec{w}_{D3}$ with themselves is $G^{ij}$, the
inner product on $\vec{w}_{F1}$ or $\vec{w}_{D1}$ is $G_{ij}$,
and cross-terms are contracted with $\delta_i^j$.

The corresponding formula for the D1/D5 system is derived by the 
S-duality transformation that takes $g_s\to g_s^{-1}$; maps the metric
$G_{ij}\to g_s^{-1} G_{ij}$ (so that $v_4\to g^{-2}_s v_4$); interchanges
the background charges 
$f_{1}\leftrightarrow d_{1}~,~n_{5}\leftrightarrow d_{5}$; 
transforms the $U(1)$ charges as 
$\vec{w}_{D1}\leftrightarrow\vec{w}_{F1}$; 
finally $\vec{p}$ ($\vec{w}_{D3}$) are (pseudo) scalars. 
This gives
\be
E R_5 = {1\over 2d_{1}}~g_s~(\vec{p} + {1\over g_s}\vec{w}_{D1})^2
 + {1\over 2d_{5}}~{g_s\over v_4}~(\vec{w}_{F1} -
{v_4\over g_s}\vec{w}_{D3})^2\ .
\label{eq:veced}
\ee
The energy formulae (\ref{eq:vecef}-\ref{eq:veced}) are derived without
assumptions about the moduli. However, in the scaling limit the 
fixed scalar conditions must be imposed.

The next step is to take the parity-odd fields into account. The vector
charges shift according to the substitution rules
\bea
p_i\to {\tilde p}_i &=& p_i + B_{ij} w^j_{F1}
- C_{ij} w^j_{D1} - A_4 w_i^{D3} - {1\over 4}
(B_{jk}C_{mi} - C_{jk}B_{mi})\epsilon^{jkml}w^{D3}_l
\nonumber\\
w_i^{D3}\to {\tilde w}_i^{D3}  &=& w_i^{D3} \nonumber\\
w^i_{F1}\to {\tilde w}^i_{F1}&=& w^i_{F1} - \chi  w^i_{D1} + 
{1\over 2}\epsilon^{ijkl}C_{jk} w_l^{D3} 
\label{eq:vshift4} \\
w^i_{D1}\to {\tilde w}^i_{D1}&=& w^i_{D1} + 
{1\over 2}\epsilon^{ijkl}B_{jk}w_l^{D3} \ .\nonumber
\eea
In these formulae there are no terms that are quadratic in the parity-odd
fields because none of the branes carrying vector charge wrap the entire
$T^4$.

The parity-odd fields also shift the background as indicated in 
(\ref{eq:bshift}). In general, this lead to significant complications; 
in particular, it is no longer consistent to choose
the background as one of the canonical ones (D1/D5 or F1/NS5),
because the parity-odd fields mix these with each other and with the
D3-branes. However, {\it when the scalars take their fixed point values} 
the result is much simpler: the shifts in background
charges are compensated by changes in the fixed scalar values.
Thus the background charges are effectively unchanged by the
parity-odd moduli. The end result for the energy of the charges
in the D1/D5 background is therefore simply (\ref{eq:veced}) 
with the shifts \pref{eq:vshift4}
and subject to the fixed scalar equations
(\ref{eq:dfixed1}-\ref{eq:dfixed3}). The corresponding result for the 
F1/NS5 system is (\ref{eq:vecef}) with the same shifts but with the fixed 
scalar equations (\ref{eq:ffixed1}-\ref{eq:ffixed3}).

\section{Entropy and U(1) Charges\label{entropy}}

In this section we take the ``heavy'' background
to consist of $n_5$ NS5-branes and $f_1$ fundamental strings.
General excitations in this 
background are characterized by the value of ``momentum'' (the scalar 
charge), energy, and the $16$ vector charges.  For sufficiently
large coupling, these general configurations are interpreted in spacetime 
as charged black holes. At small values of the charges and sufficiently 
small coupling, one has an ensemble of brane bound states in 
$AdS_3\times S^3\times T^4$. This section outlines the calculation of 
the entropy of such objects.

Consider a point in moduli space characterized by the $16$
components $G_{ij}+B_{ij}$; {\it i.e.} the RR bacground is
set to zero. The 8 NS-NS $U(1)$ currents form a 
$U(1)^4_R\times U(1)^4_L$ lattice. 
The charges are represented as vectors in the 
fashion familiar from perturbative string theory. The right and left
lattices have norms
\bea
Q_R^2 &=& (p_i + B_{ij}w^i_{F1})G^{ij} (p_j + B_{jk}w^k_{F1}) + 
w^i_{F1}G_{ij}w^j_{F1} + 2p_i w^i_{F1}\\
Q_L^2 &=& (p_i - B_{ij}w^i_{F1})G^{ij} (p_j - B_{jk}w^k_{F1})
 + w^i_{F1}G_{ij}w^j_{F1} - 2p_i w^i_{F1}\ .
\eea
The invariant norm of the full lattice is
\be
{1\over 2}(Q_R^2 - Q_L^2) = 2 p_i  w^i_{F1}\ .
\ee
Similarly, the 8 R-R $U(1)$ currents form a $U(1)^4_R\times U(1)^4_L$ 
lattice with norms given by 
\bea
P_R^2 &=& \frac1{v_4} (w^i_{D1} + {^*B}^{ij}w_i^{D3})G_{ij} 
(w^j_{D1} + {^*B}^{jk}w_k^{D3}) + 
v_4 w_i^{D3}G^{ij}w_j^{D3} + 2w_i^{D3} w^i_{D1}\\
P_L^2 &=& \frac1{v_4} (w^i_{D1} -
{^*B}^{ij}w_i^{D3})G_{ij} (w^j_{D1} - {^*B}^{jk}w_k^{D3})
 + v_4 w_i^{D3}G^{ij}w_j^{D3} - 2w_i^{D3} w^i_{D1}\ ,
\eea
and the invariant norm 
\be
{1\over 2}(P_R^2 - P_L^2) = 2 w_i^{D3}  w^i_{D1}\ .
\ee

The conformal dimensions of the theory are written
as
\be
h_{L} = {1\over 2}(\epsilon + p_5)\quad,\qquad
h_{R} = {1\over 2}(\epsilon - p_5)\quad,
\ee
where $\epsilon$ is the dimensionless  energy (energy measured in
units of the compactified dimension) and $p_5$ is the scalar charge.
These equations express the physical interpretation of conformal
levels as left and right moving energy. Vertex operators in the conformal 
field theory can be written in the factorized form 
\be
{\cal V} = {\cal V}_{\rm irr} ~ {\cal V}_{\rm U(1)} \ ,
\ee
where the latter factor carries all the dependence on the
$U(1)$ charges. In the subspace of moduli space considered here, it is 
meaningful to assert that the currents associated
to the NS-NS sublattice have level $q_1$ and those associated
to the R-R sublattice have level $q_5$. The conformal dimensions 
of the uncharged part of the operator then become\footnote{The 
normalization ${1\over 4}$ in the
conformal dimension for the $U(1)$ currents may appear unfamiliar. This 
is because we use units of $l_s=\sqrt{\alpha^\prime}$, as is standard 
in duality discussions; however, one often takes $\alpha^\prime = 2$ 
in perturbative string theory.}
\bea
h_{L}^{\rm irr} &=& {1\over 2}(\epsilon + p_5) - 
{1\over 4f_1}Q^2_L - {1\over 4n_5}P^2_L\\
h_{R}^{\rm irr} &=& 
{1\over 2}(\epsilon - p_5) - {1\over 4f_1}Q^2_R - {1\over 4n_5}P^2_R\ .
\eea
These equations express the fact that some of the excitation energy
must be expended on excitations that carry the $16$ vector charges.

The background conformal field theory is the familiar one, with central 
charge $c_R=c_L= 6f_1 n_5$. This leads to the entropy
\be
S = 2\pi\Bigl(\sqrt{c_L h_L^{\rm irr}/6}+\sqrt{c_R h_R^{\rm irr}/6}\Bigr)\ .
\ee
This is a prediction for the area of a black hole with the prescribed
set of charges. It would be cumbersome to find such a general solution 
explicitly but perhaps the result could be inferred by duality arguments
on the supergravity side.

However, in the extreme limit $h_R^{\rm irr}=0$ the area formula
is known~\cite{Dijkgraaf:1997cv}. In this case the excitation energy 
can be expressed in terms of the charges as
\be
\epsilon = p + {1\over 2f_1}Q^2_R + {1\over 2n_5}P^2_R\ .
\ee
The remaining nontrivial energy level can therefore be expressed as
\be
h_L = p + {1\over 4f_1}(Q^2_R - Q^2_L)+{1\over 4n_5}(P^2_R - P^2_L)
= p + {1\over f_1} p_i  w^i_{F1} + {1\over n_5} w_i^{D3}  w^i_{D1}\ ,
\ee
and the entropy becomes
\be
S = 2\pi\sqrt{J_3}\ ,
\ee
where
\be
J_3 = p_5 f_1 n_5 +  n_5 p_i  w^i_{F1}+ f_1 w_i^{D3}  w^i_{D1}\ .
\ee
This agrees with the area formula given in~\cite{Dijkgraaf:1997cv} upon 
specializing the $E_{6(6)}$ invariant $J_3$ to the case considered 
here~\footnote{In~\cite{Dijkgraaf:1997cv} there is also a microscopic
counting; however, the details seem different from those given 
here.}.

A special case is $p_5=0$, \ie\ no momentum flowing
on the ``large'' circle of the $T^5$. This is interesting because now
there is only momentum (and winding) along dimensions that are often
thought of as dormant; the counting argument nevertheless works.
In another special case, when the RR-charges vanish, an $SO(5)$ invariant 
formula is obtained. However, this symmetry is not 
preserved by the formalism in intermediate
steps. 

The entropy of the fundamental string with winding and momentum on 
$T^4$ is analogous to the usual Dabholkar-Harvey spectrum
of perturbative string theory, apart from the factor of $n_5$
that seems to be a ubiquitous feature when five-branes are present.

\section{A fundamental domain for the moduli space\label{globalident}}

The local geometry of the moduli space of the D1-D5 system is 
$\KK=SO(5,4)/SO(5)\times SO(4)$, the subspace of the scalar
field space $SO(5,5)/SO(5)\times SO(5)$ of supergravity on $T^4$
to which the geometry is attracted in the near-horizon limit.
The global identifications of this moduli space are implemented by 
the elements of $SO(5,5;\Z)$ which preserve the background charge vector;  
these form a subgroup $\hq \leq SO(5,5;\Z)$.
Elements of $SO(5,5;\Z)$ not in $\hq$ can be used to map D1- and D5-brane 
charges $(d_1,d_5)$ to the canonical charge vector $(N=d_1d_5,1)$, at 
the expense of mapping the moduli to exotic 
values.  In other words, one can either impose conditions such as $g_6<1$ 
and vary the background charges $(d_1,d_5)$; or allow general $g_6$, \etc, 
with $d_5=1$. We will choose the latter option. Thus, the fundamental
domain $\FF=\hq\backslash\KK$ of the moduli space consists
of several copies of the domain 
$\FF_0=SO(5,4;\Z)\backslash SO(5,4)/SO(5)\times SO(4)$,
and we will roughly be able 
to associate particular copies of $\FF_0\subset \FF$
with particular values of $(d_1,d_5)$.

The structure of the global identifications on moduli space 
$\hq\leq SO(5,5;\Z)$ is illuminated by considering particular 
$SO(2,2;\Z)=\sltwoz_L\times \sltwoz_R$ subgroups of the $SO(5,5;\Z)$ 
U-duality group of compactification on $T^4$. Such subgroups are large
enough to accomplish the transformation of the background charges to 
the canonical charge vector, and have a simple matrix realization.
A convenient choice involves $\sltwoz_R$ acting on $\tau=\chi+i\gs^{-1}$
(\ie\ the usual $\sltwoz$ symmetry of type IIB in D=10); and the
$\sltwoz_L$ acting on $\taut=A_4+iv_4 g^{-1}_s$ generated by 
$T_{1234}ST_{1234}$ and integer shifts of $A_4$. 
It is easy to check that these two $\sltwoz$'s commute. In the 
explicit computations, we will only track the dependence on these 
moduli; however, the general case is discussed at the end.

Let us represent the relevant charges by the matrix
\be
  Q=\pmatrix{f_1&d_1\cr -d_5&n_5}~,
\ee
and encode the moduli in the matrices
\bea
  \TT&=&\frac1{\tau_2}\pmatrix{1& -\tau_1\cr -\tau_1&|\tau|^2}~,\\
  & &\nonumber \\
  \Ttil&=&\frac1{\taut_2}\pmatrix{1& -\taut_1\cr -\taut_1&|\taut|^2}~,
\eea
If $g_R\in\sltwoz_R$ transforms $\tau$, and $g_L\in\sltwoz_L$ 
transforms $\taut$, then the above matrices transform as
\bea
  Q \rightarrow g_L Q g_R^t\quad&\quad 
  \TT\rightarrow (g_R^{-1})^t \TT g_R^{-1}\quad&\quad
  \Ttil\rightarrow (g_L^{-1})^t \Ttil g_L^{-1}~.
\eea
For example, the BPS tension formula for the quartet of charges in $Q$
can be written in terms of SO(2,2) invariants as
\be
  {{\tilde G}_6 M^2\over R^2_5}=\tr[\Ttil Q\TT Q^t]+2\det[Q]\ .
\label{tension}
\ee
The fixed scalars of the near-horizon limit are obtained by varying 
this expression with respect to the real and imaginary parts of $\tau$ 
and $\taut$; at the extremum two of the four parameters will be fixed.  
For instance, if there are only D-brane charges, then $A_4=\chi d_1/d_5$, 
$v_4/g_s=d_1/d_5 \gs$; in other words, $\taut=\tau d_1/d_5$. 
If we have only NS charges, then $\taut\tau=-f_1/n_5$. These relations 
are special cases of the fixed scalar conditions 
(\ref{eq:dfixed2}-\ref{eq:dfixed3}) 
and (\ref{eq:ffixed2}-\ref{eq:ffixed3}), respectively. They show how the 
moduli space can be parametrized after fixing the scalars by a single 
complex parameter that can be chosen as $\tau$.

The $SO(2,2;\Z)$ group we have isolated generally changes the
values of the brane charges, keeping only the inner product
$\det[Q]$ invariant. Among the transformations are some that can be 
used to map D1-D5 charges $(d_1,d_5)$ to the canonical background 
$(d_1d_5,1)$, while maintaining $f_1=n_5=0$; they are
\bea
  g_L&=&\pmatrix{a\,d_5&-b\,d_1\cr -c&d}~,
\nonumber\\
  g_R&=&\pmatrix{a&b\cr c\,d_1&d\,d_5}\qquad,\qquad ad\,d_5-bc\,d_1=1\ .
\label{canontransf}
\eea
Solutions to this equation exist when (and only when) $d_1$ and $d_5$ 
are mutually prime; they can be chosen to have $c=d=1$. Some freedom 
remains, because the matrices
\bea
  {\hat g}_L&=&\pmatrix{\alpha&-\beta\, d_1d_5 \cr -\gamma&\delta}~,
\nonumber\\
  {\hat g}_R&=&\pmatrix{\alpha&\beta\cr\gamma\, d_1d_5&\delta}
	\qquad,\qquad \alpha\delta-\beta\gamma d_1d_5=1~,
\label{furthertransf}
\eea
preserve the charge vector $(d_1d_5,1)$, and can be used after
the canonical transformation \pref{canontransf} to do further
shifts. Equation \pref{furthertransf} defines the congruence 
subgroup $\Gamma_0(d_1d_5)$ of $\sltwoz$. The transformations 
\pref{canontransf} that satisfy $c=d=1$ are unique for relatively prime 
$(d_1,d_5)$, modulo the $\Gamma_0(d_1d_5)$ transformations 
\pref{furthertransf}. This shows that $(d_1,d_5)$ appear precisely
once as the image of the canonical charge $(d_1d_5,1)$ under
$\sltwoz$ transformations modulo $\Gamma_0(d_1d_5)$.

\begin{figure}[p]
\epsfxsize=14cm \centerline{\leavevmode \epsfbox{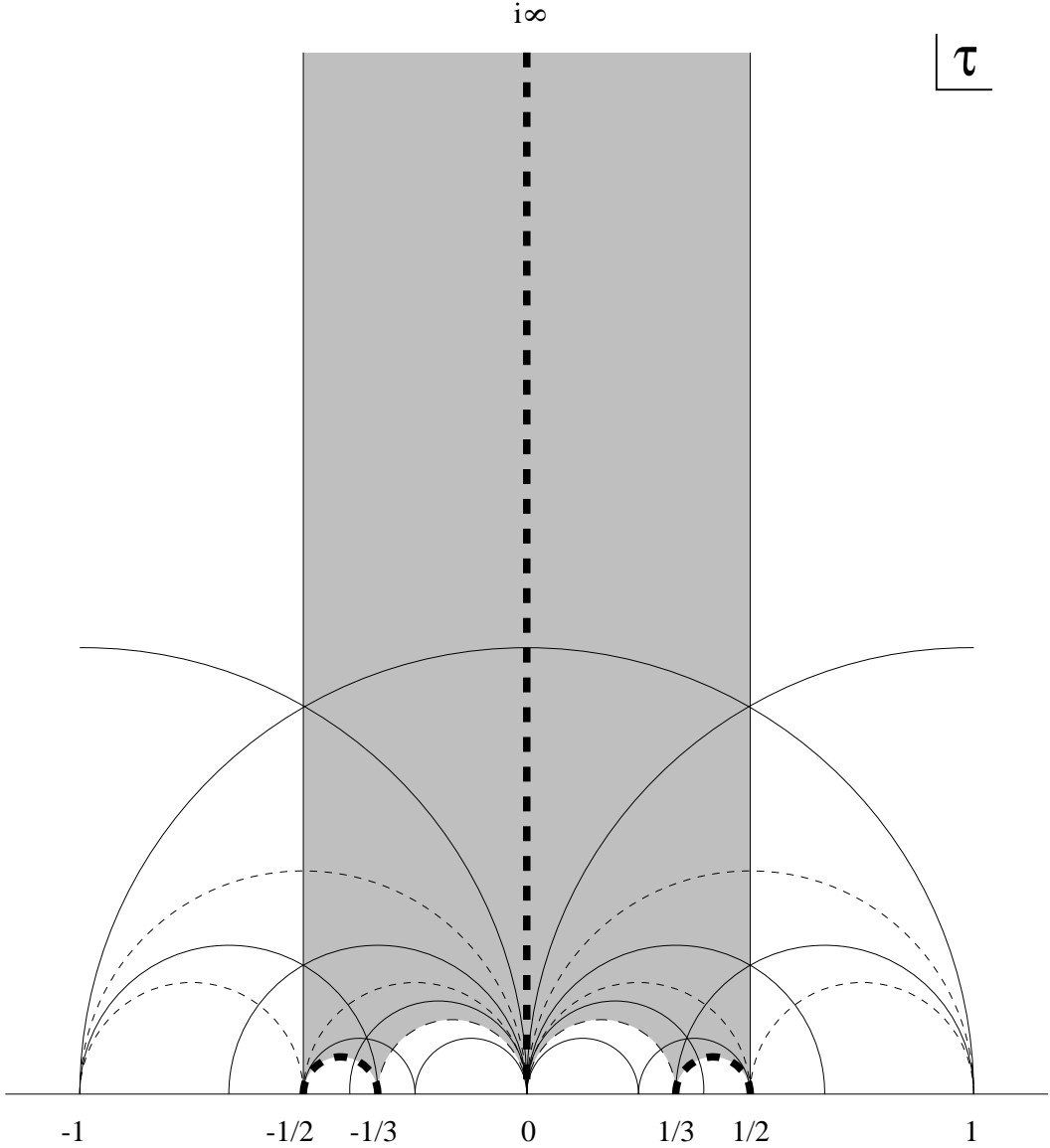}}
\caption{{\small \sl 
Fundamental domain in $\tau=\chi+i/\gs$ of the D1-D5 CFT for
$N=6$.  Corresponding boundary segments are identified under 
elements of $\Gamma_0(6)$ (in this example, the identified
segments are symmetric about the imaginary axis).
Heavy dashed lines denote the subspaces where the spacetime CFT 
becomes singular.The line from $\tau=0$ to $i\infty$ is the singular locus
for $(d_1,d_5)=(1,6)$ or $(6,1)$, with the lower end corresponding
to the former charge assignment and the upper end corresponding
to the latter.  Similarly, the circular arc from $\tau=1/3$
to $1/2$ is the singular locus for charges $(3,2)$ and $(2,3)$,
respectively.  Each rational cusp corresponds to a different
weak-coupling limit of the spacetime CFT.
}}
\label{fundomain}
\end{figure}

Now consider the fundamental domain of the residual transformations 
\pref{furthertransf}; this will be the main ingredient of a specification
of a fundamental domain for the moduli space of the spacetime CFT. Let us 
again start with non-canonical brane charges $(d_1,d_5)$. The singular 
locus, where the Coulomb branch is degenerate with the Higgs branch and 
we expect some sort of pathology in the CFT, is $\chi=0$; in other words 
$\tau$ is purely imaginary. We can use \pref{canontransf} to map the 
singular domain $\chi=0$ for charges $(d_1,d_5)$ into the fundamental 
domain for the canonical charges $(d_1d_5,1)$; this takes the modulus 
$\tau=it$ to 
\bea
  \tau'_2&=&\frac{t}{d_1^2t^2+d_5^2}~,
\nonumber\\
  \tau'_1&=&\frac{ad_1t^2+bd_5}{d_1^2t^2+d_5^2}~,
\label{singloc}
\eea
where $a$, $b$ are the parameters of the transformation \pref{canontransf} 
(with $c=d=1$).\footnote{Note that the residual transformations 
\pref{furthertransf} include elements that shifts $\chi$ by integers 
(keeping $\gs$ fixed), so if $\tau'_1$ is greater than one, 
we can shift it back into the fundamental domain.} 
The parameter $t$ is the single remaining real 
modulus of the four in $\tau$, $\taut$ --- two have
been eliminated by going to the near-horizon limit, and
one by going to the singular locus of the CFT.%
\footnote{The string coupling $\gsix$ ($\gsix^{-2}=\tau_2\taut_2$)
determined from \pref{singloc} is always larger than one after we have 
done the transformation; there does not appear to be any residual 
transformation that changes this. Charge assignments other than the
canonical one therefore always correspond to strongly coupled regions
of moduli space, from the viewpoint of the canonical background.}
Since the map of $\tau$ is in $\sltwoz$, the singular line 
$\Re\tau=0$ for charges $(d_1,d_5)$
is mapped under \pref{canontransf}
to a semicircle centered on the real axis; its
endpoints are at $b/d_5$ and $a/d_1$.
As explained in the introduction,
the singular loci for various $d_1$, $d_5$ with 
fixed $N=d_1d_5$ have physically distinct spacetime interpretations
as the places in moduli space where the spacetime CFT can
break up into its constituent D-branes;
for instance, $d_1$ `fractional instantons' might recombine
into a D-string that leaves the system by moving
onto the near-horizon Coulomb branch.
Therefore the different semicircular arcs for 
various $d_1$, $d_5=N/d_1$ correspond to 
different parts of the fundamental domain that are not identified by
the duality group.  Indeed, they are not identified under any
action of $\Gamma_0(N)$.  A picture of the fundamental domain
for $N=6$ is shown in Figure \ref{fundomain}.

Thus, the residual transformations preserving the D1-D5
charge vector $(N,1)$ and lying in the particular $SO(2,2;\Z)$
subgroup of U-duality is a certain `diagonal' $\Gamma_0(N)$.
The index of $\Gamma_0(N)$ in $\Gamma=\sltwoz$ (the number of copies of 
the $\sltwoz$ fundamental domain in a fundamental domain of 
$\Gamma_0(N)$) is~\cite{schoeneberg}
\be
 (\Gamma :\Gamma_0(N)) =  N\prod_{p|N\atop p~{\rm prime}}(1+p^{-1})~.
\ee
The number of inequivalent rational cusps in the fundamental domain
(\ie\ $\sltwoz$ images of $\tau=i\infty$ that are not equivalent
under $\Gamma_0(N)$) is~\cite{schoeneberg} 
\be
  \sigma_\infty (N)=\sum_{q|N}\varphi({\rm gcd}(q,N/q))\ ,
\ee
where $\varphi(n)$ is the Euler function
\be
  \varphi(n)=n\prod_{p|n\atop p~{\rm prime}}(1-p^{-1})\ .
\ee
We specialize to the case where the prime decomposition of $N$
contains no prime factor with multiplicity more than one;
this guarantees that any partition of $N=d_1d_5$ into $(d_1=q,d_5=N/q)$
is such that $d_1$ and $d_5$ are relatively prime, so that
the brane system is confined to the Higgs branch for
generic moduli (the extension to general $N$ is certainly of
interest but lies outside the scope of the present work).
In this case $\sigma_\infty(N)$ is simply the number of divisors
of $N$, \ie\ precisely the number of ways of partitioning $N$
into two factors $(d_1,d_5)$.  The cusps of the fundamental
domain are simply those that contain the two ends of the arcs \pref{singloc} 
for the various partitions $(d_1,d_5)$. The charges $(d_1,d_5)$ and 
$(d_5,d_1)$ are interchanged by the transformation 
$\tau\rightarrow-1/\tau$ and $\taut\rightarrow-1/\taut$.  
Hence the segment $1<t<\infty$ of the arc \pref{singloc} can be assigned 
to the former charges, and the segment $0<t<1$ to the latter charges. 
Moreover, $d_5$ copies of the $\sltwoz$ fundamental domain meet at
the cusp at $t=\infty$ along this arc, and $d_1$ copies
meet at the $t=0$ cusp at the other end of the arc
(see Figure \ref{fundomain});
this reflects the periodicity of $\chi$ in the respective original
U-duality frames, before the map to canonical charges $(N,1)$.
The cusps are the `large volume limits' of the torus 
$v_4/g_s \rightarrow\infty$, before the transformation
\pref{canontransf} to the canonical frame, for 
different pairs of charges $(d_1=q,d_5=N/q)$;
we see there is precisely one for each charge assignment.
Each large volume limit is a different weak coupling limit
for the dual CFT.  The different limits are connected through
regions of strong coupling.
The fundamental domain of $\Gamma_0(N)$ contains
all of the structure needed to describe the different
D1-D5 bound states in the context of a single connected moduli space.

To summarize, we have completely and explicitly
characterized the moduli space and its singular locus.
With only the moduli $g_s$, $v_4/g_s$, $\chi$,
and $A_4$ activated, the singular locus is given by the arcs \pref{singloc};
we can then switch on all the other moduli while preserving the fixed 
scalar conditions (\ref{eq:dfixed1}-\ref{eq:dfixed3}).
Each arc in the $\tau$-plane fundamental domain
of $\Gamma_0(N)$ gives a single connected component of the singular locus,
of codimension four in the full moduli space.
In particular, there are only as many disconnected components 
of the singular locus as there are partitions of $N$ 
into unordered pairs $(d_1,d_5)$
(for the class of examples considered here).  
All weak-coupling regions of the CFT other
than that for $d_1=N$ occur along the real axis of $\tau$;
these will be small-volume limits in the target space of
the symmetric product (see section \ref{symprod}).

Up to this point we have focussed on the moduli $\tau=\chi+i/\gs$ and 
$\taut=A_4+iv_4/\gs$ in the map to the canonical background,
by taking $B^+=0$. However, the disjoint (disconnected) nature of the 
singular loci for different $(d_1,d_5)$ is a property of generic moduli.
The simplest argument leading to this conclusion proceeds in the
F1/NS5 duality frame. We consider a singular locus which is a 
transformation to the canonical frame of the point
$\chi=C^+=0$ in the background $(f_1,n_5)$;
and another singular locus
which is the analogous transformation to the canonical frame from
the background $(f_1^\prime,n_5^\prime)$ 
(with $f_1 n_5 = f_1^\prime n_5^\prime$
and $f_1^\prime\neq n_5$). We want to determine whether these two 
co-dimension four surfaces intersect. Assume that they do, and pull 
some point on the intersection back to the original frame using the maps 
corresponding to the two inequivalent interpretations; this gives two 
points on the original singular surface corresponding to the charges 
$(f_1,n_5)$ and $(f_1^\prime,n_5^\prime)$. This surface has $\chi=C^+=0$ 
so, after taking the fixed scalar conditions in to account, it
is parametrized by the standard moduli $G_{ij}+B_{ij}$ (and $g_6$); it
is left invariant only by standard T-duality transformations.
However, this is a contradiction: $(f_1,n_5)$ and $(f_1^\prime,n_5^\prime)$ 
are not equivalent under T-duality. We conclude that the singular
surfaces do not intersect.

The case of general moduli can also be analyzed from a different
point of view. After the map to the canonical frame with charges 
$(d_1d_5,1)$, the group $\hq$ preserving the background tensor charge 
vector $\vec q$ is generated by two canonical subgroups:
(i) the further $\Gamma_0(N)$
transformations \pref{furthertransf}; 
and 
(ii) transformations $ST_\alpha S$, where $S$ is type IIB S-duality,
and $T_\alpha$ is an element of the $SO(4,4;\Z)$ T-duality group.  
In other words, any element can be described via its action on the vectors
of the charge lattice, and we can use T-duality to rotate a given
vector into some sublattice acted on
by an $SO(2,2;\Z)$ subgroup that mixes it with $(d_1,d_5)$.
The allowed transformations in this $SO(2,2;\Z)$ subgroup are the
image of $\Gamma_0(N)$ under the T-duality map.

The puzzles concerning the ``level'' of the current algebras, discussed 
heuristically in section \ref{scalimit}, can be explained in detail
using the picture developed in this section. The point is that the
concept of level is meaningful only on the singular loci, where the 
simple formula (\ref{eq:er0}) for the energy of vector excitations 
applies. Even with this restriction, the values of the levels are 
ambiguous: they are {\it covariant} under $SO(5,4;\Z)$ transformations 
so their values ($q_1$ and $q_5$) are those of the specific singular 
locus. More importantly, since the different singular loci are 
continuously related through the bulk of moduli space the levels are 
indeterminate at generic points in moduli space.

The above analysis of the global structure of the D1-D5 moduli
space carries over without significant modification to the
case of compactification on K3; in the generators
of $\hq$ discussed above, one simply replaces the $SO(4,4;\Z)$
T-duality group in transformations of type (ii) by the $SO(4,20;\Z)$
T-duality group of K3.

\section{Comparison with the symmetric orbifold CFT\label{symprod}}

The analysis of the U(1) mass formula gives a great
deal of robust, nontopological data about the spacetime CFT.
This data can be used to pin down the relation between
the symmetric orbifold CFT ${\sl Sym}^N(X^4)$ and
the spacetime CFT dual to $AdS_3\times S^3\times X^4$.
The points of comparison of this CFT with the data of
the spacetime CFT of the D1-D5 system consist of 
\begin{itemize}
\item 
The spectrum of BPS states;
\item
The qualitative density of states in the vicinity of
the orbifold locus;
\item
The moduli-dependent $U(1)^{16}$ mass formula.
\end{itemize}
In this section, we analyze these data in turn.

First, we review the spectrum of half-BPS states of the 
symmetric orbifold conformal field theory; 
it matches several previous investigations of this 
spectrum~\cite{Vafa:1996zh,Maldacena:1998bw,Larsen:1998xm,deBoer:1998ip}.%
\footnote{The discussion corrects several errors in~\cite{Martinec:1998st}.}
As argued by Vafa~\cite{Vafa:1996zh}
the half-BPS spectrum of the D1-D5 system is expected to match that
of a fundamental string with winding and momentum charges
(the so-called Dabholkar-Harvey spectrum).
This relation is most easily seen~\cite{Martinec:1999sa}
via the chain of dualities involved
in maintaining a proper low-energy description of the near-horizon
region as one passes to the core of the geometry;
these dualities map D1/D5 charge to momentum/fundamental string
winding charge.
In addition, we will find a direct map between the BPS chiral vertex
operators of the symmetric orbifold, and those constructed
in a perturbative string approach~\cite{Giveon:1998ns,Kutasov:1998zh}; 
the latter, however, do not cover the full BPS spectrum, as explained
in~\cite{Seiberg:1999xz}.  The precise match of the symmetric
orbifold BPS spectrum (including the cutoff in R-charge
at $j=d_1d_5/2$) limits one's ability to tinker with this CFT
and still match the structure of the unadulterated D1-D5 system.

The BPS spectrum is topological data, and as such does not help
us discover where the symmetric orbifold lies in the moduli space
analyzed in previous sections.  The fact that it contains 
a $U(1)^4_L\times U(1)^4_R$ current algebra of level $N=d_1d_5$
suggests it is somewhere in the $\tau=i\infty$ cusp of the fundamental
domain (\cf\ Figure \ref{fundomain}), according to the discussion
of section \ref{globalident}.
Further support for this idea
comes from consideration of the non-BPS spectrum.
An estimate~\cite{Banks:1998dd} 
of this spectrum yields Hagedorn growth $S\propto E$
in the neighborhood of the orbifold locus.  This estimate is also
consistent with $q_5=1$.

Finally, we come to the spectrum of $U(1)^{16}$ charged states.
We find a slight modification of the symmetric orbifold which precisely
agrees with the formula for vector masses at $\chi=\half$ and $B=0$, 
and identify the perturbations away from the 
orbifold point in the moduli space.

\subsection{The spectrum, BPS and otherwise}

The symmetric orbifold 
whose target space is ${\sl Sym}^N(X^4)$, where $X^4$ is K3 or $T^4$,
is an $\NN=(4,4)$ conformal field theory\footnote{The analysis in this
section partially overlaps with~\cite{Jevicki:1998rr}.}.
In this CFT, there is a huge list of BPS states;
the ground state in each twisted sector will
be in a short multiplet.  There is an independent twisted sector for each 
conjugacy class in the orbifold group, \ie\ in the present case
for each type of word in the symmetric group.
Words are composed of cycles, each of which is a cyclic permutation
of copies of $X^4$.  The twist operator for a single
cycle is the analogue of a single-trace operator 
in the 3+1d superYang-Mills/$AdS_5\times S^5$ duality --
in other words, it is the representation of a single-particle 
operator.%
\footnote{More precisely, it is the leading term in such an
operator; there could be a nontrivial, nonlinear map between the boundary
operator algebra of supergravity and that of the CFT, similar
to that seen in solvable models~\cite{Moore:1991ir}.
Indeed, we will argue below that the modulus of the CFT corresponding
to the RR scalar has a component which is not a single-particle operator
as defined here.}
Twist operators for words that are products of
cycles are multiparticle operators; their structure is determined
combinatorically, and obeys the upper cutoff $j\leq N/2$ of the allowed
R-charges (since a word in $S_N$ always has length less than 
or equal to $N$).
Thus we need only consider the twist operators for
cycles, namely the twists corresponding to
\be
  (X^4)^k/\Z_k\ .
\ee
The twist acts by cyclic permutation of the coordinates 
$Y_{a\adot}^{\sst (\ell)}$, $\lambda_{a\alpha}^{\sst (\ell)}$, 
$\bar\lambda_{a\dot\alpha}^{\sst (\ell)}$ ($\ell=0,\dots,k-1$)
of the component CFT's; the coordinates
\be
  \Ytil^{\sst(\ellt)}=
	\sum_{\ell=0}^{k-1} \exp\Bigl[2\pi i\frac{\ellt\ell}{k}\Bigr]
		Y^{\sst(\ell)}
	\quad,\qquad \ellt=0,\dots,k-1
\ee
diagonalize the action of the twist  
(similarly $\lamt$, $\bar\lamt$ diagonalize twists 
for $\lambda$, $\bar\lambda$).  The operator which creates
the twist ground state from the SL(2)-invariant CFT vacuum
is the product of the $(\ellt/k)$ twist operators 
for each $\Ytil^{\sst(\ellt)}$, $\ellt=1,...,k-1$ (and their
fermionic partners).  These component twist operators have 
conformal dimension $\Delta=\ellt(k-\ellt)/k^2$ from the 
operator that twists the bosons, and $\Delta=(\ellt/k)^2$
from the fermions; combining all the twists, the total dimension is 
\be
  \Delta_{\Z_k~\rm twist}=\hf(k-1)~.
\label{eq:totdim}
\ee
Since the dimension equals the R-charge, the operator is BPS.
We will denote the chiral part of the
resulting operator $\Sigma^{\alpha_1\cdots\alpha_{k-1}}$,
exhibiting its left-handed SU(2) R-symmetry transformation 
as a spin $\hf(k-1)$ representation.
The right-handed chiral vertex operator is similarly
$\bar\Sigma_{\dot\alpha_1\cdots\dot\alpha_{k-1}}$.

\paragraph{Details for $X^4=T^4$:} 
The highest weight $\NN=(4,4)$ vertex operators 
that create single-particle BPS states are
then built out of the twist operator $\Sigma\bar\Sigma$,
together with the diagonal fermions 
$\lamt^{\sst(0)}$, $\bar\lamt^{\sst(0)}$ which are invariant
under the twist; one simply uses these ingredients to
build operators whose dimension is equal to their SU(2) spin.  
Bosonic operators are
\bea
  \Phi_{0,0}&=&
	\Sigma^{\alpha_1\cdots\alpha_{n}}\cdot
	\bar\Sigma_{\dot\alpha_1\cdots\dot\alpha_{n}}
\nonumber\\
  \Phi_{1,1}^{ab}&=&
	\lamt_{\sst(0)}^{a(\alpha_1}\Sigma^{\alpha_2\cdots\alpha_{n})}\cdot
	\bar\lamt^{\sst(0)\,b}_{(\dot\alpha_1}
	\bar\Sigma_{\dot\alpha_2\cdots\dot\alpha_{n})}
\nonumber\\
  \Phi_{2,0}&=&
  	J_{\sst(0)}^{(\alpha_1\alpha_2}\Sigma^{\alpha_3\cdots\alpha_{n+2})}
	\cdot
	\bar\Sigma_{\dot\alpha_1\cdots\dot\alpha_{n}}
\label{bosop}\\
  \Phi_{0,2}&=&
  	\Sigma^{\alpha_1\cdots\alpha_{n}}\cdot
  	{\bar J}^{\sst(0)}_{(\dot\alpha_1\dot\alpha_2}
	\bar\Sigma_{\dot\alpha_3\cdots\dot\alpha_{n+2})}
\nonumber\\
  \Phi_{2,2}&=&
  	J_{\sst(0)}^{(\alpha_1\alpha_2}\Sigma^{\alpha_3\cdots\alpha_{n+2})}
	\cdot
  	{\bar J}^{\sst(0)}_{(\dot\alpha_1\dot\alpha_2}
	\bar\Sigma_{\dot\alpha_3\cdots\dot\alpha_{n+2})}~,
\nonumber
\eea
where $J_{\sst(0)}^{\alpha\beta}=\lamt_{\sst(0)}^{a\alpha}
\lamt_{\sst(0)}^{b\beta}\epsilon_{ab}$ is a current built
out of the diagonal fermions and 
${\bar J}^{\sst(0)}_{\dot\alpha_1\dot\alpha_2}$ is the corresponding
antiholomorphic current. 
An analogous set of chiral building blocks for boundary operators
appears in the work of~\cite{Giveon:1998ns,Kutasov:1998zh}, 
which describes perturbative
strings near the boundary of $AdS_3$; 
the transcription of the chiral vertex operators above 
to their counterparts in the notation of~\cite{Kutasov:1998zh}
is $\Sigma\leftrightarrow \WW^-$,
$\lamt\Sigma\leftrightarrow \YY^\pm$,
and $J\Sigma\leftrightarrow \XX^+$.
Since the formalism of~\cite{Giveon:1998ns}  describes a particular regime
of the spacetime CFT~\cite{Seiberg:1999xz}, the set of
chiral operators realized there is smaller than that of the
symmetric orbifold.

Similarly, the fermionic BPS operators of the $T^4$ symmetric orbifold are
\bea
  \Psi_{1,0}^a&=&
  	\lamt_{\sst(0)}^{a(\alpha_1}\Sigma^{\alpha_2\cdots\alpha_{n+1})}
	\cdot
        \bar\Sigma_{\dot\alpha_1\cdots\dot\alpha_{n}}
\nonumber\\
  \Psi_{0,1}^a&=&
  	\Sigma^{\alpha_1\cdots\alpha_{n}}
	\cdot
        \bar\lamt^{\sst(0)a}_{(\dot\alpha_1}
        \bar\Sigma_{\dot\alpha_2\cdots\dot\alpha_{n+1})}
\nonumber\\
  \Psi_{1,2}^a&=&
	\lamt_{\sst(0)}^{a(\alpha_1}\Sigma^{\alpha_2\cdots\alpha_{n+1})}
	\cdot
  	{\bar J}^{\sst(0)}_{(\dot\alpha_1\dot\alpha_2}
	\bar\Sigma_{\dot\alpha_3\cdots\dot\alpha_{n+2})}
\label{fermop}\\
  \Psi_{2,1}^a&=&
  	J_{\sst(0)}^{(\alpha_1\alpha_2}\Sigma^{\alpha_3\cdots\alpha_{n+2})}
	\cdot
        \bar\lamt^{\sst(0)a}_{(\dot\alpha_1}
        \bar\Sigma_{\dot\alpha_2\cdots\dot\alpha_{n+1})}~.
\nonumber
\eea
The sixteen operators $\Phi_{p,q}$, $\Psi_{p,q}$
fill out the Hodge diamond $H^{p,q}$ of 
$T^4$~\cite{Maldacena:1998bw,Larsen:1998xm,deBoer:1998ip}, 
with eight odd elements
corresponding to fermions and eight even elements 
corresponding to bosons.  

Finally, the proper description of the CFT for $X^4=T^4$ includes
an additional four-torus $\extrat^4$ current algebra.  
The BPS spectrum of this extra $\extrat^4$ consists of
$1$, ${\tilde\lambda}$, $\bar{\tilde\lambda}$, 
${\tilde\lambda}\bar{\tilde\lambda}$;
the counting of these latter states is
again that of the four-torus cohomology, and is isomorphic to the 
set of chiral ground states of the superstring.  
Without this extra contribution, one would not reproduce the
correct degeneracy of BPS states.
The counting of multiparticle BPS states
matches that of the chiral spectrum of the superstring, since the
basic operators \pref{bosop},\pref{fermop} 
are isomorphic to the chiral oscillator
spectrum of the superstring, and the multiparticle states are
the Fock space of the single-particle spectrum.%
\footnote{For a given element of $S_N$,
each component $T^4$ belongs to {\it some} cycle
(possibly trivial) of the permutation, hence to some oscillator 
creation operator; the total number of component $T^4$'s 
defines the total oscillator level of the chiral oscillator spectrum.}
All told, the BPS spectrum
is that of the (Dabholkar-Harvey) BPS states of a single superstring --
the 16 BPS states from the extra $T^4$ supplies the degeneracy of the
left-moving ground states, and the 
symmetric product gives the same counting as the left-moving
oscillator states at level $N=d_1d_5$;
the global $\NN=(4,4)$ 
descendants supply an extra factor of 16, representing
the right-moving ground state degeneracy.

\paragraph{Details for $X^4=K3$:} 
The situation for $X^4=K3$ is similar.  One has exactly the
same twist operators $\Sigma\cdot\bar\Sigma$ for $k$-cycles, 
which can be tensored with any of the 24 chiral primaries
from the diagonal copy of K3 in $K3^k/\Z_k$; thus the
structure is that of the chiral oscillators 
of the bosonic string.
In this case all $d_1d_5+1$ copies
of K3 are involved in the symmetric orbifold.  There is no
extra degeneracy of BPS states from some other component of the
CFT, analogous to the extra $T^4$ in the previous example;
this matches onto the uniqueness of the ground state of
the bosonic string.  Also, the additional copy of K3 in the
symmetric product means that the degeneracy of BPS states
will be that of the bosonic string at level $N=d_1d_5+1$,
matching the shift by one in ground state energy.
Finally, taking into account the 16-fold degeneracy of
the $\NN=(4,4)$ short multiplet gives the Dabholkar-Harvey
spectrum of the heterotic string, as 
expected~\cite{Vafa:1996zh,Martinec:1999sa}. The moduli involve 
the usual 20 hypermultiplets of the base K3,
together with the further blowup hypermultiplet (described below
for the torus case); together they parametrize the
$SO(4,21)/SO(4)\times SO(21)$ Teichmuller moduli space.

\paragraph{Non BPS states:} 
A simple argument~\cite{Banks:1998dd} shows that the full density of 
(generically non-BPS) NS-sector states
of the symmetric orbifold is of a stringy nature, not characteristic
of low-energy supergravity on $AdS_3\times S^3\times X^4$.
The \kth\ twisted sector has oscillators with moding $1/k$; 
the energy cost to reach this sector is $R_5E\sim O(k)$ 
(see (\ref{eq:totdim})).
Thus the total density of states is approximately
\be
  \rho(E)\sim\sum_{k=1}^{\min{(d_1d_5,R_5E)}}\rho_k\,
  \exp\Bigl[\beta_0\sqrt{k(R_5E-k)}\Bigr]~,
\label{denst}
\ee
for some constants $\beta_0$, $\rho_k$.
For a given $R_5E<d_1d_5$, the value of $k$ that dominates
the density of states is $O(R_5E/2)$,
thus $\rho(E)\sim\exp[\beta_0 R_5E/2]$ is a Hagedorn spectrum.
This spectrum turns over to the $\exp[2\pi\sqrt{d_1d_5E}]$
growth required of a CFT with central charge $c=6d_1d_5$
above the characteristic energy scale $R_5E\sim d_1d_5$.
Because the low-energy spectrum has string-theoretic
rather than field-theoretic growth,
the symmetric orbifold must describe a regime where there is a
string in the theory whose tension is of order the AdS radius
of curvature, so that the spacetime does not have a 
supergravity interpretation.  It is not difficult to
identify the culprit: In the D1-D5 background, the AdS radius
in units of the D-string 
tension is $\sqrt{d_5}$,\footnote{This is just the S-dual of a 
corresponding statement in F1-NS5 backgrounds, where the 
AdS radius in units of the F1-string scale is $\sqrt{n_5}$.}
so that when $d_5=1$ the spacetime seen by this object has
stringy curvature.  One has to make sure that the spacetime
is of low curvature with respect to all strings present in the spectrum
in order to have a valid low-energy supergravity description.
As is standard in decoupling limits, the region of the moduli
space corresponding to spacetimes with a valid low-energy
supergravity description is a strongly-coupled region
of the dual quantum field theory.
We take the qualitative agreement of the full spectrum
of the symmetric orbifold with that expected
for $d_5=1$ as further evidence for the identification
of the orbifold locus with a part of the $d_5=1$
region of the moduli space of the spacetime CFT.

\subsection{Dependence on moduli}

The $U(1)^4_L\times U(1)^4_R$ of the symmetric orbifold
$\symn(T^4)$ is at level $N=d_1d_5$,
while that of $\extrat^4$ is at level one.
This suggests that this CFT describes some part of the corresponding
region of the spacetime CFT, where the U(1) spectrum separates
into two pieces with corresponding levels --- in other words,
the $\tau\rightarrow i\infty$ cusp of the fundamental domain
in the canonical presentation of section \ref{globalident}.  We now
provide further evidence in support of
this proposal, by checking that the dependence
on all the moduli is consistent with this identification.

The moduli space of the $\symn(T^4)$ theory includes the 16 Narain moduli
\be
  \MM^{ij}=\gamma^i_{a\dot a}\gamma^j_{b\dot b}\d Y^{a\dot a}
	\bar\d Y^{b\dot b}~,
\ee
of the $d_1d_5$ copies of $T^4$ in the symmetric product, as
well as the four blowup modes of the $\Z_2$ twist in $S_{d_1d_5}$;
near the orbifold point, the corresponding operators are the descendants 
\be
  \MM^{ab}=G_{-1/2}^{a\alpha}\bar G_{-1/2}^{b\dot\beta}
	\Sigma_\alpha\cdot\bar\Sigma_{\dot\beta}~,
\label{twistmod}
\ee
of the $\Z_2$ twist highest weight.
In addition to these 20 moduli parametrizing
$\KK=SO(5,4)/SO(5)\times SO(4)$,
there are the U(1) current perturbations 
$\JJ_i\overline \JJ_j$,
$\Jtil_i\overline \JJ_j$,
$\JJ_i\Jtilbar_j$, and
$\Jtil_i\Jtilbar_j$,
where $\JJ_i$, $\overline\JJ_i$ are the eight diagonal U(1) currents 
of $(T^4)^N$; and $\Jtil_i$, $\Jtilbar_i$ are the U(1) currents
of the $\extrat^4$.
One issue we will have to address is precisely how the 
moduli space of the spacetime theory sits inside this 84-dimensional
moduli space.

To begin, let us turn off the self-dual NS B-field moduli
in the U(1) spectrum formula (\ref{eq:veced}) 
(with the shifts (\ref{eq:vshift4}) implied).
Let $d_1=N$ and $d_5=1$, and define the unit volume metric
$\ghat_{ij}=G_{ij}/\sqrt{v_4}$ and six-dimensional string coupling
$\gsix=\gs/\sqrt{v_4}$.  The U(1) mass formula reduces to
\bea
  R_5 M &=&\frac{1}{2N}\Bigl[
   (p_i-C_{ik}w^k_{D1}+N\chi w^{D3}_i)\cdot \gsix\ghat^{ij}\cdot
   (p_j-C_{jm}w^m_{D1}+N\chi w^{D3}_j) \nonumber \\
	& &\qquad\qquad+
	  w_{D1}^i\cdot\gsix^{-1}\ghat_{ij}\cdot w_{D1}^j + 
     2p_i w^i_{D1}\Bigr] \nonumber \\ 
    & &+ \half\Bigl[
   (w_{F1}^i-{^*C}^{ik}w^{D3}_k-\chi w_{D1}^i)\cdot \gsix\ghat_{ij}\cdot
   (w_{F1}^j-{^*C}^{jm}w^{D3}_m-\chi w_{D1}^j)  \nonumber \\
	& &\qquad\qquad+
	  w^{D3}_i\cdot\gsix^{-1}\ghat^{ij}\cdot w^{D3}_j
	- 2 w^i_{F1} w_i^{D3}\Bigr]~.
\label{modcompare}
\eea
The $SO(4,4)/SO(4)\times SO(4)$ moduli parametrized by $\gsix$, $\ghat$, 
and $C_{ij}$ appear differently in the two square brackets.  This is 
because the eight U(1) charges $(p_i,w_{D1}^i)$ transform in the vector
of the associated $SO(4,4;\Z)$ T-duality group, while
the eight charges $(w^i_{F1},w^{D3}_i)$ transform in a spinor representation
(the two are related by triality in SO(4,4)).
In terms of this data, we identify the moduli of the tori
in the symmetric orbifold as a metric $\gsix^{-1}\ghat_{ij}$
and an antisymmetric tensor $C_{ij}$, whereas the moduli
of the extra $\extrat^4$ are a metric $\gsix^{-1}\ghat^{ij}$
and an antisymmetric tensor ${^*C}^{ij}=\epsilon^{ijkl}C_{kj}$.
Note also that, even though $w_{F1}^i$ has a contravariant index, 
it should be thought of as a `momentum' charge on $\extrat^4$
for the present purpose, since it is this quantity
that is shifted by the $\extrat^4$ `winding' $w^{D3}_i$ 
in the presence of a nontrivial antisymmetric tensor background.

Having found the embedding of the $SO(4,4)/SO(4)\times SO(4)$
moduli of the spacetime CFT in that of the $(SO(4,4)/SO(4)\times SO(4))^2$
moduli space of tori of the ${\sl Sym}^N(T^4)\times\extrat^4$ CFT,
we can account for the $\chi$ dependence as follows.  From the
way it appears in \pref{modcompare}, part of
the perturbation corresponding to 
$\chi$ should be another antisymmetric tensor modulus
of ${\sl Sym}^N(T^4)\times\extrat^4$ since it couples in
the same way; however, since it couples the momentum
of ${\sl Sym}^N(T^4)$ to the winding on $\extrat^4$
and vice-versa, it must involve an antisymmetric perturbation
that mixes the two factors.  
The fact that the orbifold is parity symmetric means that it
can only describe $\chi=0$ or $\chi=\half$ in the moduli 
space~\cite{Witten:1997yu}. However, $\chi=0$ is a singular CFT, leaving 
$\chi=\half$ as the only candidate.
Thus we should modify the orbifold to include an asymmetric shift
by half a lattice vector coupling $\JJ_i$ and $\Jtil_i$.
Note that this asymmetric shift will copy the winding $w^{D3}_i$
on the $\extrat^4$ onto {\it each} copy of $T^4$ in the symmetric
product, accounting for the relative factor of $N$ in shift of 
the momenta in the first term of \pref{modcompare}.  The asymmetric shift
amounts to turning on a term
\be
  \widehat\MM=\Bigl(\JJ^{a\dot a}\Jtilbar^{b\dot b}
	-\Jtil^{a\dot a}\overline\JJ^{b\dot b}\Bigr)
	\epsilon_{ab}\epsilon_{\dot a\dot b}~,
\ee
in the action of the sigma model on ${\sl Sym}^N(T^4)\times\extrat^4$.
However, the vertex operator corresponding to the modulus $\chi$
cannot consist solely of $\widehat\MM$, since deforming
the orbifold to $\chi=0$ would not result in a singular CFT.
This job is accomplished by 
the scalar $\Z_2$ blowup mode $\epsilon_{ab}\MM^{ab}$
\pref{twistmod}, which also has the right quantum numbers 
to be part of the modulus $\chi$.  Moreover, it is the operator that
shifts the theta angle in the symmetric orbifold sigma model;
turning it on by half a unit tunes the theta angle to zero,
resulting in the singular CFT expected at $\chi=0$.  
Thus the dependence of the spacetime
CFT on the modulus $\chi$ is carried by several different 
parts of the symmetric orbifold -- the blowup modulus
accounts for the singularity encountered at $\chi=0$,
while the current-current perturbation $\widehat\MM$ takes care
of the mixing of U(1) charges as a function of $\chi$.
The correct perturbation away from the orbifold point is
$\delta\chi\cdot(\widehat\MM+\epsilon_{ab}\MM^{ab})$.

The analysis for the other
three blowup moduli (self-dual NS two-forms in the present
duality frame) is similar, but complicated by
the fact that these perturbations enter as metric deformations
in the U(1) mass formula.  Indeed, they couple
the `momenta' $p_i$ and $w_{F1}^i$ to one another, as
well as the `winding' quantum numbers $w_{D1}^i$ and $w^{D3}_i$.
It is natural to identify the
perturbations away from the orbifold point with 
the symmetric combinations 
\be
  \vec\MM=\Bigl(\JJ^{a\dot a}\Jtilbar^{b\dot b}
	+\Jtil^{a\dot a}\overline\JJ^{b\dot b}\Bigr)
	\vec\tau_{ab}
	\epsilon_{\dot a\dot b}~.
\ee
It is not too hard to check that, to lowest order, the effect
on the symmetric product is $N$ times that on the extra $\extrat^4$,
just as for the $\chi$ perturbation.  In any case, these moduli
are turned off at the orbifold locus (otherwise, they would 
break various discrete symmetries).  The perturbation
away from the orbifold locus is
$\delta {\vec B^+}\cdot(\vec\MM+{\vec\tau}_{ab}\MM^{ab})$.

To summarize, we propose that
the orbifold ${\sl Sym}^N(T^4)\sdtimes\extrat^4$,
where the semi-direct product is meant to indicate the
additional asymmetric shift coupling the two factors,
describes the line $\chi=\half$ in the D1-D5 moduli space
in the canonical duality frame of section \ref{globalident}.
Moving in the moduli subspace parametrized by
$\chi$ and $\gsix=\gs/\sqrt{N}$, we should encounter 
the singularities corresponding to all other partitions of $N$.
These loci depend separately on $d_1$ and $d_5$, and we 
get a different singular locus 
in the moduli space of the symmetric product,
for each possible partition of
the product $N=d_1d_5$ into two factors. 
All the other singularities are at small volume of the $T^4$,
with order one amount of the blowup modulus turned on;
thus the sigma model is strongly coupled.

It would be interesting to understand what phenomenon in the
CFT distinguishes the singular domains; it is natural
to speculate that there is
a singularity in the sigma model target space in which
$d_5$ `fractional instantons' come together to make
a D1-brane or D5-brane that leaves the system, and that 
this phenomenon is related to the denominators of 
the rational cusps of the fundamental domain --
\ie\ that $\chi\sim a/d_1$ or $\chi\sim b/d_5$ (in the notation
of section (\ref{globalident})).
Perhaps these are relics of the $d_1$- or $d_5$-twisted
sectors of the orbifold, since the `fractional instantons'
at the orbifold point are the fields parametrizing the
component copies of $T^4$; and it is these twist operators
that sew together the proper number of `fractional instantons' 
of the orbifold theory to reconstitute a string that can
leave the system.  It is hard to be precise, however,
since the orbifold point is far in the moduli space
from the singular loci.

\subsection{Other orbifolds}

{\bf Adding KK monopoles:}
In~\cite{Kutasov:1998zh}, the GKS formalism was extended to the $\NN=(4,0)$
SCFT obtained when one introduces Kaluza-Klein monopoles
into the brane background, after compactification on an
additional circle.  The moduli space of the IIB theory on $T^5$
is $E_{6(6)}/USp(8)$, which is restricted to $F_{4(4)}/SU(2)\times USp(6)$
by the fixed scalar mechanism.
Analogous to the D1-D5 system without the KK monopole, the 
group of discrete identifications of the moduli space
is the subgroup of $E_{6(6)}(\Z)$ that fixes the background charges.
It would be interesting to give a more explicit characterization
of it, along the lines of the present discussion for the
D1-D5 background.
In the symmetric orbifold CFT, the addition of $p$ KK monopoles
to the background involves an asymmetric orbifold that acts 
as a $\Z_p$ twist of the right-moving SU(2) 
R-symmetry~\cite{Kutasov:1998zh,Sugawara:1999qp}.  
This twist breaks the
right-moving supersymmetry completely, consequently there will be
additional moduli which involve lowest components of superfields
from the untwisted sector (one can check that no new moduli arise
from twisted sectors of the $\Z_p$ orbifold).
Near the orbifold point, the additional moduli are generated
by the vertex operators 
\be
  \MM_{a\adot\,5}=\d Y_{a\adot} \bar J_3~,
\ee
(here $J_3$
is the R-current left unbroken by the $\Z_p$ action), and by
\be
  \widetilde \MM^{ab}=G_{-1/2}^{a\alpha}\Sigma_\alpha\cdot
  	\bar\lamt_{\sst(0)}^{b\dot\beta}\bar\Sigma^{\dot\alpha}
		\tau^3_{\dot\alpha\dot\beta}~.
\label{KKtwistmod}
\ee
It is not difficult to check that these operators have 
no three-point function with themselves, the $T^4$ moduli,
or the blowup modes \pref{twistmod}, and are thus moduli
(this should be a sufficient condition given $(4,0)$ supersymmetry).
They also have the right spacetime quantum numbers 
under both Lorentz and $SU(2)\times USp(6)$ transformations.
Specifically, the moduli transform as the $(2,14)$ under this group,
which decomposes as a $(2,1,4)\oplus(2,2,5)$ under
the manifest $SU(2)\times SU(2)\times SO(5)$ of the orbifold CFT
(the $SU(2)\times SU(2)$ transforms the left-movers of the CFT,
the SO(5) the right-movers); the (2,2,5) consist of the moduli
from the untwisted sector, and the (2,1,4) are the eight moduli
\pref{twistmod},\pref{KKtwistmod}
coming from the $\Z_2$ twisted sector of the symmetric orbifold.

One can again carry out the scaling analysis of section~\ref{scalimit}.
The type IIB theory is now compactified on $T^6$, with one
circle kept large and the rest having sizes
of order the string scale.  There is a background of three
of the 27 available tensor charges turned on, which we have taken to
consist of $p$ KK monopoles, $q_5$ D5-branes, and $q_1$ D-strings.
The background naturally distinguishes one of the five small circles
as the nontrivial fiber of the KK monopole;
the remaining four are isomorphic to the $T^4$
that the D1-D5 system is compactified on.
The vector excitations about the background form a ${\bf 27}$
of the $E_{6(6)}$ duality group of the small $T^5$:
${5\choose 3}=10$ D3-branes, 5 D1-branes, 5 fundamental strings,
5 KK-momenta, one wrapped D5-brane, and one wrapped NS5-brane.
These naturally decompose as $\bf 8_v+8_s+8_c+1+1+1$ 
under the $SO(4,4)$ T-duality group of the $T^4$.
One finds a structure reminiscent of \pref{eq:er0}, with three separate
$T^4$ current algebras of levels $q_1$, $q_5$, and $p$;
and an interesting triality symmetry among the terms.
We hope to give further details elsewhere.

{\bf Orientifolds:}
One can also consider the D1-D5 system in the presence of an
orientifold plane.  In this case, the orientifold projection
eliminates the RR scalar and the four-form $A_4$, as well as
the NS B-field.  Therefore, the $SO(5,4)/SO(5)\times SO(4)$
near-horizon moduli space will be reduced to $SO(4,4)/SO(4)\times SO(4)$;
the RR scalar will be frozen to either $\chi=0$ or $\chi=\half$
for any given values of $d_1$ and $d_5$.
In particular, the moduli that enabled us to move between the various
domains associated to different one-brane and five-brane charges
for fixed $N=d_1d_5$ are projected out.
Any orbifold conformal field theory that describes such a situation
then has no analogues of the twisted sector moduli $\MM^{ab}$;
the orbifold moduli space will then only describe $d_1=1$ or $d_5=1$.

A general analysis of D5-branes in the presence of orientifold 5-planes
was performed in~\cite{Witten:1998kz,Hori:1998iv,Gimon:1998be}.  
There are two types of orientifold
that fix $\chi=0$ and two that fix $\chi=\half$; only the latter
are expected to lead to nonsingular dual CFT's when D1-branes
are added and the scaling limit is taken.  These are
\begin{itemize}
\item
An O5-plane of SO-type (D5-brane charge -1) with an odd number of 5-branes,
yielding gauge group $SO(2k+1)$;
\item
An O5-plane of Sp-type (D5-brane charge +2)
with an even number of D5-branes, giving $Sp(2k)$ gauge group.
\end{itemize}

One might therefore expect that there is an appropriate modification 
of the symmetric orbifold that generates the moduli space of
a single $SO(2k+1)$ or $Sp(2k)$ instanton on $T^4$ or K3.
The dimension of the moduli space of
$n$ instantons in $SO(2k+1)$ or $Sp(2k)$
gauge theory on $T^4$ or K3 is $4[2k(n-k)-k]$, leading one
to expect a CFT of central charge $c=6[2d_1d_5-d_5]$,
in particular one has $c=6d_5$ when $d_1=1$.
The orientifold will eliminate all half-integer spin representations
of R-symmetry from the half-BPS spectrum.  It is this projection
that removes the twisted moduli $\MM^{ab}$ \pref{twistmod}
related to $\chi$ and $B^+$, pinning the orbifold at $\chi=\half$.
Since the projection eliminates the gauge fields that carry
fundamental string and D3-brane flux,
when $X^4=T^4$, only the U(1) charges associated to 
momenta and wrapped D1-branes survive.


\vskip 1cm

{{\Large \bf Acknowledgments:}}
We wish to thank 
A. Dabholkar,
R. Dijkgraaf,
J. Harvey,
A. Hashimoto,
T. Hollowood,
J. Maldacena,
A. Strominger,
and
particularly D. Kutasov 
for helpful discussions; and 
the Harvard University theory group 
and the ITP, Santa Barbara for hospitality during the course of 
our investigations. This work was supported by DOE grant 
DE-FG02-90ER-40560.
F.L. is supported in part by a McCormick Fellowship.

\appendix
\section{Derivation of a BPS formula}
The mass formulae of interest are BPS formulae, and 
hence direct consequences of supersymmetry. BPS masses are derived by 
solving appropriate eigenvalue problems. The details of this strategy as
well as original references can be found in the 
review~\cite{Obers:1998rn}.

The simplest starting point is the expression for the supersymmetry 
algebra in M-theory. Imposing nontrivial supersymmetry leads to
an eigenvalue equation for the central charges
\be
\left[ {\cal C}\Gamma^M P_M + {1\over 2}{\cal C}\Gamma^{MN} Z_{MN}
+{1\over 5!}{\cal C}\Gamma^{MNPQR} Z_{MNPQR}\right]\epsilon = 0~.
\ee
The central charges $Z_{MN}$ and $Z_{MNPQR}$ are M2- and M5-brane charges, 
respectively, and the $P_M$ are the momenta; in particular $P_0$ is the mass 
$M$ that we want to compute. The parity-odd moduli are taken to vanish; 
they will be reinstated in the end. $\epsilon$ is the spinorial eigenvector 
of the preserved supersymmetry. The metric is mostly plus. 

In this Appendix we consider a F1/NS5 background and excitations with 
arbitrary vector charges. Our charge assignments are type IIB
but transformation to M-theory is simple; this leads to the
eigenvalue equation
\be
\left[\Gamma^5 Q_1 + \Gamma^{1234}\Gamma^5 Q_5 + \Gamma^i P_i + 
\Gamma_i \Gamma^{11}W^i_{F1}+ \Gamma_i \Gamma^{5}W^i_{D1}
+\Gamma^{5}\Gamma^{11}\Gamma^{1234}\Gamma^i W_i^{D3}\right]\epsilon = 
\Gamma^0 M \epsilon
\ee
The physical charge, denoted by a capital letter, is the mass of 
an isolated brane with the given charge. The masses (\ref{masses})
therefore give the conversion factors between the physical charges 
used here, and the quantized charges used in the main text.

The square of the eigenvalue equation gives the expression for the
mass
\be
M^2 = M^2_T + M^2_V + 2M_{TV}~,
\label{eq:tempmass}
\ee
where
\be
M^2_T = Q^2_1 + Q^2_5 + 2Q_1 Q_5 \Gamma^{1234}~,
\ee
is the contribution of the background, 
\bea
M^2_V &=& \vec{P}^2 + \vec{W}^2_{F1} + \vec{W}^2_{D1} + \vec{W}^2_{D3} 
\nonumber \\
& &\quad + 2\Bigl[ \Gamma^{11}\vec{P}\cdot\vec{W}_{F1} +  
\Gamma^{5}\vec{P}\cdot\vec{W}_{D1} \nonumber\\ 
& &\qquad\qquad -\Gamma^{11}\Gamma^{1234} \vec{W}_{D1}\cdot\vec{W}_{D3}
+\Gamma^{5}\Gamma^{1234} \vec{W}_{F1}\cdot\vec{W}_{D3}  \nonumber \\
& &\qquad\qquad\qquad +  \Gamma^5 \Gamma^{11}\Gamma_{ij}W^i_{D1}W^j_{F1}
+\Gamma^5 \Gamma^{11}\Gamma^{1234}\Gamma^{ij}W_i^{D3}P_j\Bigr]~,
\eea
is the contribution of the vector excitations, and
\bea
M_{TV} &=& Q_1 \left[ - \Gamma_i W^i_{D1} + 
\Gamma_{11}\Gamma^{1234}\Gamma^i W_i^{D3}\right]
\nonumber \\ 
&+&Q_5 \left[\Gamma^5 \Gamma^{1234}\Gamma^i P_i
- \Gamma^5 \Gamma^{11}\Gamma^{1234}\Gamma_i W^i_{F1}\right]~,
\eea
is an interaction term between the background and the excitations. In these 
equations it is implied that all gamma-matrices act on $\epsilon$, even 
though this is not indicated explicitly.

At this point the manipulations simplify upon taking the scaling limit 
into account. $M_T^2$ is order $l_s^{-4}$, $M_V^2$ is order $l_s^{-2}$, 
and $M_{TV}$ is order $l_s^{-3}$.  The square of the largest terms is
\bea
(Q_1 Q_5 \Gamma^{1234} + M_{TV})^2 &=& Q_1^2 Q_5^2 
+Q^2_5 (\vec{P}^2 + \vec{W}^2_{F1} + 2\Gamma^{11}\vec{P}\cdot\vec{W}_{F1}) 
\nonumber \\
 & &\quad 
+Q^2_1 ( \vec{W}^2_{D1} + \vec{W}^2_{D3}- 2\Gamma^{11}\Gamma^{1234} 
\vec{W}_{D1}\cdot\vec{W}_{D3}) \\
& &\quad
+ 2Q_1 Q_5 \left(-\Gamma^5 \Gamma^{1234}\vec{P}\cdot\vec{W}_{D1}
-\Gamma^5 \vec{W}_{F1}\cdot\vec{W}_{D3} \right. \nonumber \\
& &\quad\qquad
\left. \Gamma^5 \Gamma^{11}\Gamma^{1234}\Gamma_{ij}W^i_{F1}W^j_{D1}
+ \Gamma^5 \Gamma^{11}\Gamma^{ij}P_i W_j^{D3}\right)~, 
\nonumber
\eea
Note that $\Gamma^{1234}$ anticommutes with $M_{TV}$ so that there
are no terms of order $l_s^{-3}$. It is this structure of the supersymmetry
which is ultimately responsible for the vector charges having finite
energy in the scaling limit.
 
Next, take the square root and expand
\bea
Q_1 Q_5 \Gamma^{1234} + M_{TV}
&\simeq&  Q_1 Q_5 
+ {Q_1\over 2Q_5} ( \vec{W}^2_{D1} + \vec{W}^2_{D3}- 
2\Gamma^{11}\vec{W}_{D1}\cdot\vec{W}_{D3})
\nonumber\\
& &\quad\quad + {Q_5\over 2Q_1} ( \vec{P}^2 + \vec{W}^2_{F1}- 
2\Gamma^{11}\vec{P}\cdot\vec{W}_{F1}) 
\\
& &\quad\quad\quad - \Gamma^5 \vec{P}\cdot\vec{W}_{D1}
-\Gamma^5 \vec{W}_{F1}\cdot\vec{W}_{D3}
\nonumber\\
& &\quad\quad\quad\quad 
+ \Gamma^5 \Gamma^{11}\Gamma_{ij}W^i_{F1}W^j_{D1}
+ \Gamma^5 \Gamma^{11}\Gamma^{ij}P_i W_j^{D3}\nonumber~.
\eea
This result is valid up to terms of order $l_s^{2}$. We have chosen 
the subspace where $\Gamma^{1234}=1$, as appropriate when $Q_1,Q_5>0$; 
this is legitimate at this point because $\Gamma^{1234}$ commutes with 
all other operators that remain. The terms in (\ref{eq:tempmass}) now 
add to 
\bea
M^2 &=& (Q_1 + Q_5)^2  +
{Q_1+Q_5\over Q_5} ( \vec{W}^2_{D1} + \vec{W}^2_{D3}- 2\Gamma^{11}
\vec{W}_{D1}\cdot\vec{W}_{D3}) \nonumber \\
& &\quad +{Q_1+Q_5\over Q_1}  ( \vec{P}^2 + \vec{W}^2_{F1} + 
2\Gamma^{11}\vec{P}\cdot\vec{W}_{F1})~, 
\label{eq:gamma11}
\eea
up to order $l_s^{2}$. Note that this simple expression is the
result of numerous cancellations between the vector terms and the 
tensor-vector interactions. At this point
the eigenvalue equation is solved by taking $\Gamma^{11}=1$ (this 
is legitimate because it commutes with $\Gamma^{1234}=1$). Taking
the square root again gives
\bea
M &=& Q_1 + Q_5 + 
{1\over 2Q_5} ( \vec{W}^2_{D1} + \vec{W}^2_{D3}- 2
\vec{W}_{D1}\cdot\vec{W}_{D3}) \nonumber \\
& &\quad
+ {1\over 2Q_1}  ( \vec{P}^2 + \vec{W}^2_{F1}- 2\vec{P}\cdot\vec{W}_{F1})~. 
\eea
The excitation energy can now be read off as the
energy above the background mass. After conversion to
the quantized charges used in the main text, the result reads
\be
E R_5 = {1\over 2f_1}~(\vec{p} + \vec{w}_{F1})^2
 + {1\over 2n_5}~{1\over v_4}~(\vec{w}_{D1} - v_4\vec{w}_{D3})^2~.
\label{eq:veced2}
\ee
Here it is understood that $G^{ij}$ (contracting $p_i$, $w_i^{D3}$) and 
$G_{ij}$ (contracting $w^i_{F1}$, $w^i_{D1}$) are measured in units
of $l_s$. The final equation (\ref{eq:veced2}) is identical to 
(\ref{eq:veced}), except that the sign of $\vec{w}_{D3}$ has been changed. 
This is purely a matter of conventions.

Two refinements of these results are needed for the entropy computation
in section \ref{entropy}. First, it is straightforward to generalize the 
computation in this Appendix by including also the scalar charge $p_5$, 
{\it i.e.} the momentum along the background $F1$. The result is simply 
that $p_5$ can be added on the right-hand side of (\ref{eq:veced2}), 
without any further cross-terms. Next, the mass computed here is
the largest eigenvalue of the central charge matrix. The computation 
also identifies the smaller eigenvalue: it is found by choosing
instead $\Gamma^{11}=-1$ after (\ref{eq:gamma11}). This results in
the opposite signs within the absolute squares in (\ref{eq:veced2}). This
justifies the smaller conformal weight ($h_L$) used in section 
\ref{entropy}.


\end{document}